\documentstyle[12pt,aaspp4]{article}
\accepted{}
\slugcomment{Submitted for publication to The Astrophysical Journal Supplement}
\lefthead{Eskridge et al.}
\righthead{IR vs.~Optical Morphology}

\received{2002 March 4}
\begin{document}
 
\title{Near-IR and Optical Morphology of Spiral Galaxies\footnotemark}

\author{Paul B.~Eskridge\altaffilmark{2,3}, Jay A.~Frogel\altaffilmark{2,4,5,6},
Richard W.~Pogge\altaffilmark{2,4}, Alice C.~Quillen\altaffilmark{4,7}, Andreas
A.~Berlind\altaffilmark{8}, Roger L.~Davies\altaffilmark{9}, 
D.L.~DePoy\altaffilmark{2,4}, Karoline M.~Gilbert\altaffilmark{2}, Mark 
L.~Houdashelt\altaffilmark{10}, Leslie E.~Kuchinski\altaffilmark{4,11}, Solange 
V.~Ram\'{\i}rez\altaffilmark{4,12}, K.~Sellgren\altaffilmark{2,4}, Amelia 
Stutz\altaffilmark{2}, Donald M.~Terndrup\altaffilmark{2,4} \& Glenn 
P.~Tiede\altaffilmark{4,13}}

\footnotetext{Based partially on observations obtained at the Cerro Tololo
Interamerican Observatory, operated by the Association of Universities for
Research in Astronomy, Inc.~(AURA) under cooperative agreement with the
National Science Foundation}
\altaffiltext{2}{Department of  Astronomy, The Ohio State University, Columbus,
OH 43210}
\altaffiltext{3}{Department of Physics and Astronomy, Minnesota State 
University, Mankato, MN 56001}
\altaffiltext{4}{Visiting Astronomer, Cerro Tololo Interamerican Observatory}
\altaffiltext{5}{Visiting Senior Scientist, Lawrence Berkeley National 
Laboratory, Berkeley, CA 94720}
\altaffiltext{6}{present address: NASA Headquarters, Code S, 300 E Street SW, 
Washington, DC 20546}
\altaffiltext{7}{Department of Physics and Astronomy, University of Rochester, 
Rochester, NY 14627}
\altaffiltext{8}{Department of Astronomy and Astrophysics, University of 
Chicago, Chicago, IL 60637}
\altaffiltext{9}{Department of Physics, South Road, University of Durham,
Durham DH1 3LE, England}
\altaffiltext{10}{Department of Physics and Astronomy, Johns Hopkins University,
Baltimore, MD 21218}
\altaffiltext{11}{NASA/IPAC, California Institute of Technology, MS 100-22,
Pasadena, CA 91125}
\altaffiltext{12}{Astronomy Department, California Institute of Technology, MS 
105-24, Pasadena, CA 91125}
\altaffiltext{13}{Department of Astronomy, University of Florida, Gainesville, 
FL 32601}

\authoremail{paul.eskridge@mnsu.edu}

\begin{abstract}
We announce the initial release of data from the Ohio State University Bright 
Spiral Galaxy Survey, a $BVRJHK$ imaging survey of a well-defined sample of 205
bright, nearby spiral galaxies.  We present $H$-band morphological 
classification on the Hubble sequence for the OSU Survey sample.  We compare 
the $H$-band classification to $B$-band classification from our own images and 
from standard galaxy catalogs.  Our $B$-band classifications match well with 
those of the standard catalogs.  On average, galaxies with optical 
classifications from Sa through Scd appear about one T-type earlier in the 
$H$-band than in the $B$-band, but with large scatter.  This result does not 
support recent claims made in the literature that the optical and near-IR 
morphologies of spiral galaxies are uncorrelated.  We present detailed 
descriptions of the $H$-band morphologies of our entire sample, as well as $B$- 
and $H$-band images for a set of 17 galaxies chosen as type examples, and $BRH$ 
color-composite images of six galaxies chosen to demonstrate the range in 
morphological variation as a function of wavelength.  Data from the survey are 
accessible at {\tt http://www.astronomy.ohio-state.edu/$\sim$survey/}
\end{abstract}


\section{Introduction}

Galaxy morphology has been a cornerstone of extragalactic research even before
the discovery of the extragalactic nature of the nebulae (e.g., 
\markcite{old}Parsons 1850).  Detailed morphological classification of 
galaxies, in the modern sense dates to the work of \markcite{tnf}Hubble (1936), 
who established the now classic tuning-fork diagram for galaxy morphology.  
Since then there have been a number of modifications of this basic 
classification scheme.  \markcite{hag}Sandage (1961) described the sequence in
substantially more detail than did \markcite{tnf}Hubble (1936).  
\markcite{dvd}de Vaucouleurs (1959) presents a scheme that accounts for inner 
ring and spiral structure, as well as strong (SB) and weak (SAB) bar classes.
A later version (\markcite{rc2}de Vaucouleurs, de Vaucouleurs \& Corwin 1976) 
introduces the numerical ``T-type'' coding for morphology.  This scheme is 
described in its most modern form in \markcite{rc3}de Vaucouleurs et al.~(1991, 
hereafter RC3).  \markcite{v76}van den Bergh (1976 and references therein) 
introduced the DDO luminosity classification scheme that relates the surface 
brightness and the appearance of spiral arms with the absolute luminosity of 
the galaxy.  \markcite{ee2}Elmegreen \& Elmegreen (1982) created a 
classification scheme for the degree of structure seen in the spiral arms of
galaxies, ranging from flocculent (arm class 0) to grand design (arm class 12).
\markcite{rng}Buta (1986) and \markcite{rn2}Buta (1995) added a much more 
detailed classification scheme for resonance rings following the general 
precepts of \markcite{dvd}de Vaucouleurs (1959).  The most significant 
departure from the basic Hubble scheme is that proposed by \markcite{wwm}Morgan 
(1958), and given in its final form by \markcite{mkw}Morgan, Kayser \& White 
(1975).  This scheme is essentially a one-dimensional classification based on 
the degree of concentration of light, with a secondary `form' parameter to 
distinguish amongst gross galaxy types.

A common property, and, therefore, a common weakness of all these schemes has 
been the nearly exclusive use of $B$-band plate material for galaxy 
classification (although see \markcite{fzw}Zwicky 1955 and 
\markcite{sch}Schweizer 1976 for early work on multiwavelength morphology).  
This constrains existing classification schemes in two major ways.  First, as 
classification has been done primarily in the $B$-band, it is very sensitive to 
the distribution of blue stars and dust.  This is particularly unfortunate for 
the study of late-type galaxies, as the distribution of the young, blue stars 
and the dust can be very different from the distribution of the total baryonic 
mass.  Second, although well-exposed plates, taken under good seeing 
conditions, can offer excellent spatial resolution, plates are poor photometric 
detectors, and have a very limited dynamic range.  Thus, for instance, it is 
possible to miss features such as nuclear bars in classification plates that 
are deeply exposed in order to reveal the structure of spiral arms in the outer 
disk (see \markcite{bar}Eskridge et al.~2000 for a discussion of this effect).  

The advent of two-dimensional near-infrared (near-IR) detectors has resulted in
numerous efforts (e.g., \markcite{hs}Hackwell \& Schweizer 1983; 
\markcite{hfd}Thronson et al.~1989; \markcite{bw}Block \& Wainscoat 1991; 
\markcite{bea}Block et al.~1994; \markcite{bnp}Block \& Puerari 1999) to 
compare galaxy morphology in the optical and near-IR.  These studies revealed
that there can be substantial differences between the optical and near-IR
morphologies of spiral galaxies.  Indeed, \markcite{bnp}Block \& Puerari (1999)
assert that there is no correlation between the optical and near-IR 
classifications of spirals.  The main problem with these studies is that they 
deal with single galaxies, or at most, small samples of objects.  In fact, the 
lack of a large, well-defined sample of multiwavelength digital imaging of 
spirals prompted the creation of the OSU Survey.  

We believe that classification in both the optical and near-IR, of a large, 
statistically complete sample of spirals is an important step in our 
understanding of the physical morphology of galaxies as a function of 
wavelength.  The astrophysical motivation for such a comparison is that very 
different sorts of stars dominate the emission from spiral galaxies at optical 
($B$-band) and near-IR wavelengths.  In the optical, the regions of current 
massive star-formation, typically associated with the spiral arms will 
dominate.  Further, the absorbing effect of interstellar dust is substantial in 
the optical, and this material is also typically concentrated along the spiral 
arms.  In the near-IR, the flux is dominated by light from old giants.  There 
can be significant contributions from young supergiants, but this effect 
appears to be small in all but the most extreme situations 
(\markcite{r98}Rhoads 1998).  

A number of recent studies have attempted to classify galaxies in some sort of 
automated way (e.g., \markcite{a94}Abraham et al.~1994; \markcite{oea}Odewahn 
et al.~1996; \markcite{nrg}Naim, Ratnatunga \& Griffiths 1997; 
\markcite{onp}Odewahn et al.~2002).  Such efforts, while promising, have mainly 
been devoted to classification at the very simplest level; determining if a 
given extended object is a pure spheroid, a disky system, or peculiar.  There 
have been pioneering efforts aimed at objective bar and spiral structure 
classification (e.g., \markcite{a99}Abraham et al.~1999; \markcite{onp}Odewahn 
et al.~2002), and such efforts have largely been directed at studying 
high-redshift galaxy samples.  This is appropriate as such samples are very 
large (thousands of objects), and each galaxy is typically only a few 
resolution elements in diameter.  Detailed classification of well-resolved 
galaxies is still the provenance of ``expert classifiers''.  A landmark study 
of the classification biases of different ``expert classifiers'' 
(\markcite{nea}Naim et al.~1995) reveals that different classifiers are in
general agreement with one another, but typically have a scatter of about one
T-type.  Whether this is good or bad agreement is, of course, a subjective
decision.  We believe it is good enough, given the underlying purpose of
morphological classification.  Although descriptive classification is not 
physical, it is an essential starting point for a physical classification 
scheme:  Morphology is akin to species taxonomy.  Species taxonomy is not 
genetics, but it is an essential guide to asking relevant questions in genetics 
(e.g., \markcite{bio}Mayr 1942).

In this paper we present the Ohio State University Bright Spiral Galaxy Survey 
sample; a large, statistically well-defined sample of bright nearby spiral 
galaxies.  While previous multi-object samples (most notably that of 
\markcite{fgt}Frei et al.~1996) have been a great boon for many sorts of 
projects, none of these samples cover a wavelength range from the optical 
through the near-IR, and are both large and selected according to a 
well-defined set of criteria.  We hope that the availability of the OSU Survey 
data will be of use to the community.  In \S 2 we describe the survey selection 
and data taking procedure.  We present statistical results of the comparison of 
the $B$- and $H$-band morphologies of the sample in \S 3, and give notes on the 
$H$-band morphologies in \S 4.  In \S 5 we discuss some implications of our 
results, and suggest promising areas for future research.

\section{The OSU Bright Spiral Galaxy Survey}

The Ohio State University Bright Spiral Galaxy Survey comprises deep, 
photometrically calibrated $BVRJHK$ images of 205 spiral galaxies selected from 
the \markcite{rc3}RC3 according to the following criteria:  $0 \leq T \leq 9$; 
$M_B \leq 12$; $D \leq 6\rlap.'5$; $-80^{\circ} < \delta < +50^{\circ}$ (due to 
the pointing limits of the CTIO 1.5m and Perkins 1.8m reflectors).  We have 
been unable to obtain near-IR data for twelve galaxies from the defined sample 
that are south of $\delta = -25^{\circ}$ (we have optical data for six of these 
galaxies).  Figure 1 shows histograms of \markcite{rc3}RC3 T-types for all 
galaxies in the \markcite{rc3}RC3 with $M_B \leq 12$ and \markcite{rc3}RC3
ellipticity $< 0.5$ (solid histogram), and for the OSU Sample as defined above 
(dotted histogram).  It is clear from this figure that our diameter and 
declination cut-offs do not impose any morphological selection bias on the OSU 
Survey.

We obtained the data for the survey in a large number of observing runs with
six different telescopes of apertures between 0.9m and 2.4m, in both 
hemispheres in the period from 1993 through 2000.  The $JHK$ data were obtained
mainly with the 1.8m Perkins reflector of the Lowell Observatory, and the
CTIO 1.5m, with some additional data from the 2.4M Hiltner telescope of the
MDM observatory.  We used OSIRIS (\markcite{d93}DePoy et al.~1993) at the 1.8m 
and the 1.5m.  We also used CIRIM ({\tt 
http://www.ctio.noao.edu/instruments/ir\_instruments/cirim/cirim.html}) at the 
1.5m.  The data from the 2.4m were taken with TIFKAM (\markcite{p98}Pogge et 
al.~1998).  The $BVR$ data were obtained mainly with the 1.8m Perkins reflector 
and the CTIO 0.9m, with additional data from the USNO 1.1m Hall Telescope of 
the Lowell Observatory, the 1.3m McGraw-Hill telescope of the MDM observatory, 
and the CTIO 1.5m.  A large number of imaging cameras and detectors were used 
to obtain the $BVR$ imaging.

For the NIR imaging, our typical observing strategy was as follows:  We would
obtain a series of short (1 to 20 sec) observations with no read-out or dither 
(a stack).  We would then read out the stack, dither the telescope, and take 
another stack.  We would then chop by a distance comparable to the field of
view, and take a similar set of dithered sky observations.  The total on-source 
integration time per galaxy in $J$ and $H$ are typically 10 to 15 minutes.  In 
$K$ we observed somewhat longer, generally between 20 and 30 minutes.  For the 
optical imaging, our typical strategy was to obtain 3 images per filter, with 
total integration times of 20-30 minutes in $B$, 10-15 minutes in $V$, and 5-10 
minutes in $R$.  As we had many nights of observing time that were 
non-photometric, we adopted a strategy of obtaining deep images on 
non-photometric nights, and using photometric nights to obtain short 
integrations (``snapshots'') of large numbers of galaxies to calibrate our
deep images.  Our data are unavoidably heterogeneous, having been obtained 
at many telescopes, with many different instruments, by many different 
observers, and under a wide range of sky conditions.  However, the approximate 
limiting surface brightnesses of typical Survey data are $B_{lim} \approx 26$ 
magnitudes per square arcsecond and $H_{lim} \approx 20$ magnitudes per square 
arcsecond.

Information about the OSU Survey is available at {\tt 
http://www.astronomy.ohio-state.edu/$\sim$survey/}.  With the acceptance of 
this paper for publication, we shall make the uncalibrated $B$ and $H$ images 
available as an {\it Early Data Release} at {\tt 
http://www.astronomy.ohio-state.edu/$\sim$survey/EDR/index.html}.  We plan on 
making the entire calibrated $BVRJHK$ data set publicly available when we are 
satisfied with the photometric calibration.  Anyone who is interested in using 
data that have not been publicly released yet is encouraged to contact one of 
the first three authors.

\section{$H$- and $B$-band Morphology}

In Table 1 we present the sample.  Column 1 gives the NGC, IC, ESO or Anon
designations of the galaxies in the sample.  In columns 2 and 3 we give the 
morphological types from the \markcite{rc3}RC3, and the \markcite{cag}Carnegie 
Atlas of Galaxies (Sandage \& Bedke 1994, hereafter referred to as CAG).  
Columns 4 and 5 present our $B$- and $H$-band classifications.  Our $B$-band 
classification follows the precepts laid out in the \markcite{rc3}RC3 and 
\markcite{cag}CAG, and is generally closer to those of the \markcite{rc3}RC3.  
Our $H$-band classification follows these same general precepts, but with a 
number of differences driven by the nature of the $H$-band morphology of 
spirals.  The classic prescription for the classification of spirals involves 
three observables:  the bulge to disk ratio; the pitch angle of the spiral 
arms; the degree of resolution (or knottiness) of the arms.  For two of these 
three observables, there is a clear bias toward an earlier classification from 
near-IR images.  In near-IR images, the bulge is more prominent, and the spiral 
arms are less knotty than they appear in optical images.  It is not obvious 
that the pitch angle of the spiral arms should change systematically from the 
optical to the near-IR.  \markcite{bnp}Block \& Puerari (1999) find 
substantially different pitch angles in optical and near-IR images of a small 
sample of spirals, but they do not see evidence for any systematic trend.  As 
the degree of resolution in the arms of even very late type galaxies is 
substantially lower in the $H$-band than in the optical, the essential features 
in determining the $H$-band Hubble stage are the bulge to disk ratio, and the 
pitch angle of the spiral arms.

As we discussed in \S 1, the optical and near-IR emission from spirals is
dominated by different populations of stars, and subject to different levels
of dust absorption.  Thus structures that are dominated by older stellar
populations should be more prominent in the near-IR than the optical.  This has 
the consequence that the bar fraction in the near-IR is significantly higher 
than it is in the $B$-band (\markcite{bar}Eskridge et al.~2000).  It should 
also cause the observed bulge to disk ratio to be higher in the near-IR than in 
the optical.  This should tend to drive near-IR morphological classification to 
earlier types.  Structures that are dominated by either younger stellar 
populations, or by absorption effects will be less prominent in the near-IR 
than the optical.  Flocculent spiral features, either dominated by OB stars or
by dust lanes, may disappear in the near-IR, again leading to earlier
morphological classification.  Also features such as resonance rings that are 
currently lit up by star formation may be less prominent or unseen in the 
near-IR (see the images of NGC 6782 on the Survey Web-page for an example).  
Grand design spiral patterns are seen in near-IR imaging of many spirals (e.g., 
\markcite{pea}Puerari et al.~2000), thus spiral patterns are not traced out by 
only the young stars.  In fact, there is evidence from the study of a small 
sample that there can be drastic differences in the pitch angle of the spiral 
pattern measured in the optical and the near-IR (\markcite{bnp}Block \& Puerari 
1999).  The assertion of \markcite{bnp}Block \& Puerari (1999) that the optical 
and near-IR morphologies of spiral galaxies are uncorrelated begs for a 
comparative study of the optical and near-IR morphological types of a large 
unbiased sample of spirals such as ours.

One of us (PBE) classified the entire sample in both $B$ and $H$ twice, with 
excellent overall agreement between the trials in each waveband: more than 80\% 
of the classifications agreed to within two subtypes (i.e., Sa to Sb).  We also 
cross-checked our results by comparing classifications done by two of us (PBE 
\& JAF).  We selected forty galaxies from the sample in $B$, and forty in 
$H$, and compared our classifications for these subsets.  For both the $B$ and 
$H$-band subsets, more than 80\% of the tested galaxies were assigned types 
that again agreed to within two subtypes.  In no case were there any systematic
differences between any of our sets of classifications.  This compares well 
with the scatter between classifiers reported by \markcite{nea}Naim et 
al.~(1995).  Figure 2 shows comparisons of classifications, parameterized as 
T-types, following the \markcite{rc3}RC3.  We compare the optical 
classifications of the \markcite{rc3}RC3 and the \markcite{cag}CAG to one 
another, and both of these catalog classifications with our optical and near-IR 
classifications.  Table 2 gives the average and median differences and 
dispersions as a function of Hubble type for each of the classification pairs 
shown in Figure 2.  We note that we are presently interested in Hubble stage
and bar family classification only.  Although we note the presence of inner (r)
and outer (R) ring structures when they are obvious, we otherwise feel
unqualified to make variety-classifications (rs and s types) as defined by
\markcite{dvd}de Vaucouleurs (1959), or detailed ring morphology
classifications as defined by \markcite{rng}Buta (1986; \markcite{rn2}1995).

Figure 2a and Table 2a show a good agreement between the \markcite{rc3}RC3 and 
the \markcite{cag}CAG for types Sab through Sc ($2 \leq T \leq 5$).  For 
galaxies with \markcite{rc3}RC3 types S0/a and Sa, the \markcite{cag}CAG gives 
slightly later classifications.  It is well known that the \markcite{cag}CAG 
classifies very few galaxies as later than Sc, thus the disagreement with the 
\markcite{rc3}RC3 at late types is expected.  Figure 2b and Table 2b show our 
OSU $B$-band types against the \markcite{rc3}RC3 types.  Overall, the agreement 
is excellent.  We tend to classify galaxies with \markcite{rc3}RC3 types S0/a 
and Sa ($0 \leq T \leq 1$) as having marginally later type, and galaxies with 
\markcite{rc3}RC3 types later than Sc ($T > 5$) as having marginally earlier 
types.  However, our optical classifications agree with the \markcite{rc3}RC3 
to within the dispersion for all types S0/a through Sm.  In Figure 2c and Table
2c we compare our $B$-band types with the \markcite{cag}CAG.  For galaxies with 
\markcite{cag}CAG types of S0/a through Sc ($0 \leq T \leq 5$) our optical 
types match even better than with the \markcite{rc3}RC3.  The disagreement at 
later types is consistent with what is seen in Fig.~2a.

In Figure 2d and Table 2d, we compare our $H$-band types with the 
\markcite{rc3}RC3 types.  For the earliest (S0/a and Sa, or $0 \leq T \leq 1$) 
and latest (Scd through Sm or $6 \leq T \leq 9$) types, the agreement is 
excellent.  But for galaxies with \markcite{rc3}RC3 types from Sab through Sc
($2 \leq T \leq 5$), our $H$-band types are roughly one sub-class earlier than 
the \markcite{rc3}RC3 types on average.  While this is within the dispersion 
for any given type, the effect is persistent across the range of mid-type 
spirals.  We compare our $H$-band types with the \markcite{cag}CAG in Figure 2e
and Table 2e.  Our $H$-band classifications are earlier on average by about one 
subclass from \markcite{cag}CAG types S0/a through Sbc ($0 \leq T \leq 4$).  At 
Sc ($T=5$), we are in rough agreement, and there are too few galaxies with 
\markcite{cag}CAG types later than Sc for any meaningful comparison.  Finally, 
in Figure 2f and Table 2f, we compare our $B$-band classifications with our 
$H$-band classifications.  For all OSU $B$-band types Sa through Scd ($1 \leq T
\leq 6$), we assign $H$-band types that are roughly one subclass earlier than 
the optical types.  At the very earliest (S0/a) and latest (Sd through Sm)
types, our $B$-band classifications agree with our $H$-band classifications on
average.

In summary, we find that spiral galaxies with optical classifications of 
roughly Sab through Sc ($2 \leq T \leq 5$) appear about one T-type earlier in 
the $H$-band on average.  This result persists for all optical classifications 
(\markcite{rc3}RC3, \markcite{cag}CAG, and our own).  And the disagreement 
between optical and near-IR classes occurs in the range of morphological types 
over which all optical classifications are in good agreement.  For the earliest 
(S0/a and Sa) and latest (Scd through Sm) types , there is no wavelength 
dependence in the morphological classification.  While there have been a number 
of studies comparing galaxy morphology in the optical and near-IR (see \S 1), 
these have mainly been studies of individual galaxies, or at most small 
samples.  Ours is the first result showing the nature of the morphological 
K-correction (the change in morphology as a function of observed 
rest-wavelength)for a large, statistically complete sample of spirals.  Thus it 
is worth considering possible physical reasons for our results.

For the earliest types (S0/a and Sa), there is no disagreement in the mean 
between the optical and near-IR classifications.  As these galaxies have very 
little ongoing star-formation, and very little dust, both the optical and
near-IR morphologies are dominated by the distribution of old, late-type stars.
The two wavebands should give similar classifications, and we are reassured 
that they, in fact, do so.

We find an average disagreement of one subclass between the optical and near-IR
classifications of mid-type spirals (optical classifications Sab through Sc),
in the sense that the near-IR classifications tend to be earlier than the
optical classifications.  As noted above, bulges are more prominent and spiral
arms are less knotty in the near-IR than in the optical.  This drives the
earlier classification of mid-type spirals in the near-IR.

We see no disagreement in the mean for the latest type galaxies (Scd
through Sm).  These are galaxies that are dominated by on-going star formation,
so this seems surprizing at first.  However, very late-type galaxies have
poorly defined spiral structure, and very weak or non-existent bulges as 
matters of definition.  The things that work to drive the sort of morphological
K-correction we see for the mid-type spirals are the relative prevalence of
the bulge, and the difference in contrast and appearance of spiral arms traced
by sites of on-going star formation (in the optical) or by the distribution of
old, late-type giants (in the near-IR).  Galaxies with no bulge, and very 
poorly defined spiral patterns will not look substantially different in the
near-IR than in the optical, and will thus be classified essentially the same
on average.

\section{Notes on Individual Galaxies}

Below, we provide brief descriptions of the $H$-band morphology of each
survey galaxy, along with our $H$-band morphological classifications.  For
galaxies with OSU$B$ types more than two sub-types different (earlier in all
cases) than their OSU$H$ types, we discuss the features that drive the earlier 
classification in the $H$-band.  Table 3 gives a set of $H$-band type examples 
for each morphological class.  We show $B$- and $H$-band images of these type 
examples in Figure 3.  The images in Fig.~3 are typically shown in an 8-bit
logarithmic stretch with white set at 2$\sigma$ below the sky level, and black
at the peak value.  We adjusted this in cases with bright nuclear sources.  In 
Figure 4, we show color-composite images (blue is $B$, green is $R$, red is 
$H$) of six galaxies.  We selected two early-type, two mid-type, and two 
late-type spirals to show in Figure 4.  Within each pair, we selected one 
object that is given the same classification in the optical and near-IR, and 
one that is classified at least a full Hubble type earlier in the near-IR than 
in the $B$-band.

NGC 150:
SBb:  Strong bar.  Two-arm grand design spiral.  Asymmetric arms that start
before the intersection point with the bar.  Regions of strong star formation
in the areas near the arm-bar contacts.  SE arm bifurcates just beyond the bar
contact point.  NW arm is more linear than the SE arm.

NGC 157:
Sbc (Type example -- see Fig.~3f):  No apparent bar.  Grand-design spiral
pattern.  Oval bulge with the arms beginning at the ends of the bulge {\it
minor} axis.  SW arm initially linear, then begins to wind, and bifurcate as it
crosses the disk major axis.  Arm fades after only 180 degrees.  NE arm is very
wide to NE, and appears to wrap a full turn before it fades away.  The optical
types for this galaxy are SAB(rs)bc: (\markcite{rc3}RC3), Sc(s)II-III
(\markcite{cag}CAG), and SBbc (OSU$B$).  It is thus a rare example of a galaxy
that appears optically barred, but unbarred in the near-IR.  What appears to be
a bar in the optical looks much more like the result of projection in the
near-IR.

NGC 210:
(R)SB0/a:  Strong bulge threaded by thick, high-contrast bar with visible ansae
at the ends.  Faint outer ring (not obviously a spiral pattern) with
star-forming knots.

NGC 278:
Sb:  Face-on.  Centrally condensed, resolved circular nucleus, embedded in a
slightly elliptical bulge.  Fairly high B/D ratio.  Two-armed spiral pattern,
with tightly wrapped spiral arms.  Arms have obvious dust lanes, and numerous
bright star-forming knots.

NGC 289:
SB(r)ab (Type example -- see Fig.~3d):  Strong bar and aligned elliptical
bulge.  Symmetric two-arm pattern emerges from the ends of the bar.  Arms are
smooth, and well-defined.  Arms both bifurcate after $\sim$90 degrees.  Log
stretch shows bright ansae at the ends of the bar.

NGC 428:
SBm (Type example -- see Fig.~3k):  Nuclear point-source embedded in an
asymmetric inner disk.  No true bulge.  Outer disk has numerous star-forming
knots, and evidence for two very asymmetric spiral arms.  The NW arm is fairly
tightly wrapped, and the better defined of the two in it's inner part.  It has
several very bright knots through $\sim$60 degrees of winding to the east.  It
then becomes very diffuse and poorly defined, but can still be traced as a low
surface brightness (LSB) feature through a winding angle of $\sim$180 degrees.
The SE arm is very loosely wrapped, and LSB.  It can be followed for $\sim$100
degrees to the west.

NGC 488:
Sa (Type example -- see Fig.~3c):  System nearly face-on.  Bright, circular
nucleus, embedded in a large, slightly elliptical bulge.  The bulge and disk
have the same PA.  There is a faint spiral patterns, with thin, smooth
tightly-wrapped arm features embedded in a smooth disk.  There are no prominent
star-forming knots.

NGC 578:
SBc:  Very small, slightly elliptical bulge, with prominent bar running E-W.
System at fairly high inclination.  Arms originate at ends of bar, at very
steep angles (closer to 120 degrees and 90 degrees).  Arms well-defined for
almost 180 degrees, then they break up.  SE arm bifurcates.  NW arm becomes
very diffuse.  A few knots in the inner arms.  Outer arms are very lumpy.  A
bright companion is just to the east, superposed on one bifurcation of the SE
arm.

NGC 613:
SB(r)bc (Type example -- see Fig.~3f):  Bright nuclear point-source.  Nuclear
bar aligned with elliptical bulge and large normal bar.  Fairly highly inclined
disk.  Two bright arms emerge from the ends of the bar, with many bright knots
near the bar ends.  SE arm wraps loosely to the NE.  NW arm wraps tightly to
the SW, and forms an inner-disk ring.  Several other, lower surface-brightness
arms emerge from the ring.

NGC 625:
Sm (Type example -- see Fig.~3k):  No central concentration.  Thick disk, seen
close to edge on.  No evidence of spiral structure.  System rich is
star-forming knots/bright star clusters.

NGC 685:
SBcd (Type example -- see Fig.~3h):  Bright elongated nucleus, extending into a
thin bar.  Patchy, LSB arms emerge perpendicular to the bar ends.  Arms fade
out after $\sim$270 degrees.

NGC 779:
SAB(r)b:  Bright nucleus embedded in an elliptical/boxy bulge.  System has high
inclination.  Grand-design two-armed spiral pattern, with the arms emerging
from the corners of the boxy bulge.  The inner arms are high surface brightness
(HSB), and form an inner pseudo-ring, with bright enhancements along the major
axis.  The arms are embedded in a smooth disk, with no pronounced star-forming
knots.  The break in the pseudo-ring to the NW side of the nucleus may be due
to dust extinction.  At lower surface brightness levels, the system is smooth,
and extends to very large radii.

NGC 864:
SBb:  Strong, slightly curved bar, with bright ansae at the ends.  Asymmetric
arms that start before the intersection point with the bar.  SE arm has a kinky
appearance; more like a series of line-segments than a smooth curve.  Arms make
less than a half turn before losing their integrity.  Outer spiral pattern is
flocculent.

NGC 908:
Sc:  Bright elliptical bulge, no evidence of a bar.  Three-arm spiral pattern,
with east arm much more open than north and SW arms.  May be an inner ring in
the disk.  East arm bifurcates soon after it originates from a possible inner
ring.  SW arm is also peculiar.  It emerges from the inner disk on the west
side, curves to the south, and reconnects with the inner disk near the origin
point of the east arm.  Arms are patchy, with many obvious star-forming knots.

NGC 986:
No $JHK$ data.

NGC 988:
SBcd:  System close to edge-on.  Nuclear point-source embedded in a thin,
slightly asymmetric bar.  Inner bar is HSB.  Outer bar is much lower surface
brightness.  Two thick, knotty, diffuse, filamentary spiral arms begin at the
ends of the LSB bar, and wrap for at least 180 degrees.  A bright star is
superposed on the NW side of the disk.

IC 239:
SBc:  System nearly face on.  Small circular bulge embedded in a thick bar.
Disk is very LSB, with smooth two-armed grand design spiral pattern.  Arms seem
to begin from the long end of the bar (they are very fat at their base), and
wind through $\sim$180 degrees before fading.  LSB nature of the arms makes any
internal structure hard to distinguish.

NGC 1003:
Scd?:  Nearly edge-on.  Weak nuclear points-source embedded in a very flattened
bulge/lens.  Inner part of LSB disk has same PA as bulge/lens feature.  At
larger radii there is an obvious isophotal twist/warp.  The are traces of a
flocculent spiral pattern, and several knots of star formation.

NGC 1042:
SABc:  Weak bar.  Nuclear point-source extends into mildly elliptical bulge.
Classic two-armed spiral.  Arm emerging from the NW side of the bulge has an
HSB inner portion, becomes diffuse after $\sim$30 degrees, and then becomes
well defined again after $\sim$180 degrees.  There are a number of bright
star-forming knots in the outer part of this arm.  The NW arm winds through a
total of $\sim$270 degrees before fading.  SE arm is more regular, with an
inner HSB region fading into an outer LSB region, and not re-brightening.  This
arm can also be traced through $\sim$270 degrees before fading.

NGC 1058:
Sa:  System close to face on.  Centrally peaked nuclear source.  Clear break in
surface brightness between the inner and outer disk.  Very faint, flocculent
irregular spiral pattern, with tightly wound arm segments.

NGC 1073:
SB(r)ab:  Strong, thin bar, and very elliptical bulge.  Weak grand-design arms
form an inner ring.  Arms originate at the ends of the bar, and are
roughly at right-angles to the bar.  The $B$-band type is SBc for all
classifiers.  We classify NGC 1073 substantially earlier in the $H$-band
because the outer, open spiral arms are very faint in the near-IR, leaving the
inner arm/ring structure as the spiral feature that dominates the
classification.

NGC 1084:
Sb:  Bright nuclear point-source.  Bulge elongated along disk major axis; no
sign of a bar.  Multi-arm pattern, but arms not very well defined.  Inner arms
have a number of bright knots.  Arms wind about 180 degrees before fading from
view.

NGC 1087:
SBd:  Face-on system with a small nuclear bar, and no true bulge.  Bar is
embedded in a knotty LSB disk, with wisps of spiral structure.  At low SB,
there is evidence for two very fat, open spiral features.

NGC 1187:
SB(r)b:  Strong bar emerging along the major axis of an elliptical bulge (with
nuclear point-source).  Spiral arms emerge from the ends of the bar.  Arm
beginning in the SE (and turning to the west) is fairly regular.  Arm beginning
at the NW end of the bar begins heading N, then fades out, and re-appears to
the east, heading SE.  Arms have a few bright knots.

NGC 1241:
SB(r)ab:  Nuclear point-source.  Bulge becomes elliptical/boxy at lower
intensity levels.  Prominent bar emerges from opposite corners of the boxy
bulge.  Strong two-arm pattern with arms originating in bright ansae at the
ends of the bar.  North arm is much more tightly wrapped than south arm.
Several other arms appear at low surface-brightness levels.  LSB disk extends
further to the NW than the SE (disk major axis runs SE-NW).  This may be due to
an interaction with a fainter peculiar late-type galaxy to the NE.

NGC 1255:
No $JHK$ data.

NGC 1300:
SBb (Type example -- see Fig.~3e):  Strong bar, aligned with bulge major axis.
Arms commence at the ends of the bar, and are roughly perpendicular to the bar.
Arms extend roughly 180 degrees before fading.  Far ends of the arms straighten
out, rather than curving back into a pseudo-ring.  Arms rich with star-forming
knots.

NGC 1302:
SB0:  Bright circular bulge, with a weak roughly N-S bar.  Smooth lens
outside bar.  Perhaps a faint outer ring.  No evidence for spiral structure.

NGC 1309:
SABb:  Small, bright, circular bulge; bulge becomes more elliptical at larger
radii.  No other indication of a bar.   Flocculent, multi-armed spiral pattern.
Spiral arms very patchy and knotty.  Arms stronger to the NE than the SW.  Most
prominent arm appears to continue south on the SE side of the galaxy, tracing
out a series of bright knots in a straight, due south line.  The optical types
for this galaxy are SA(s)bc: (\markcite{rc3}RC3), Sc(s)II (\markcite{cag}CAG),
and Scd (OSU$B$).  The substantially earlier type in the near-IR is due to the
enhanced prominence of the bulge in the $H$-band compared to the optical.

NGC 1317:
SB0 (Type example -- see Fig.~3a):  Nuclear point source embedded in a nuclear
bar.  Inner isophotes of the bulge are elliptical with the same major axis as
the bar, but much less flat.  At lower intensity the bulge becomes circular.
At lower intensity still, the isophotes become elliptical again, but the major
axis is roughly perpendicular to the major axis of the nuclear region.  This
appears to extend into a second, very faint, large scale bar (again roughly
perpendicular to the inner bar).  The disk is of quite low surface brightness,
and has a faint inner ring at the radius where the outer bar fades away.

NGC 1350:
(R)SB(r)a (Type example -- see Fig.~3c):  Bright nuclear point-source.
Elliptical bulge, oriented with the major axis nearly N-S.  Bar, with major
axis skewed NE-SW from bulge major axis.  Bar most obvious from bright ansae at
bar ends.  These bridge into an inner ring (or tightly wound spiral arms).
Faint, LSB spiral features emerge from the ring at the ends of the apparent
major axis, and wrap to form an outer ring.

NGC 1371:
SAB(r)a (Type example -- see Fig.~3c):  Bright, centrally condensed bulge.
Elliptical, and oriented NW-SE at high light-levels.  At lower light-levels, a
bar is apparent, and the isophotes twist clockwise.  Low surface-brightness
arms emerge from ansae at the bar ends.  Pitch angle fairly steep.  No obvious
knots in arms.  Arms appear to wrap around to form a faint outer ring.

NGC 1385:
SBdm (Type example -- see Fig.~3j):  Faint, small bulge, with stubby nuclear
bar threading it.  Patchy, flocculent spiral arms emerge from several locations
in the inner disk.  The inner arms, especially are full of star-forming knots.
Inner arms include two straight segments heading just east of north from the
bar area.  Outer disk is asymmetric, with very little emission on the NW side
of the galaxy.

NGC 1421:
Sc:  System close to edge-on.  Bright symmetric nucleus embedded in an
elliptical bulge.  There appears to be a diffuse bar with a PA offset $\sim$30
degrees from the bulge PA.  Two spiral arms emerge from the bar feature.  The
arms are very well defined, but the pattern is quite asymmetric.  The disk PA
is $\sim$0 degrees.  The south arm emerges from the bar, then bends slightly to
the south, and goes south in a straight line almost all the way to the edge of
the disk.  It is thin, HSB, and well defined.  It has several strong
knots/clumps of star formation along it.  The arm bends to the east at the
southern edge of the disk, and quickly fades.  The south side of the disk (east
of the arm) is knotty, and shows signs for one or two lower surface brightness
armlets.  The northern arm bends smoothly to the west from the bar, then
bifurcates.  The inner (western) segment is thick, and has a number of bright
knots, before terminating abruptly at a very elongated bright feature.  The
outer segment is smooth, and LSB.  It extends approximately as far to the north
as the southern arm does south, but is much less well defined.

NGC 1425:
No data.

NGC 1433:
No $JHK$ data

NGC 1493:
SBb:  Nuclear point-source embedded in elliptical bulge with a small, high
contrast bar.  Flocculent spiral pattern emerges, with structure apparent
starting at the ends of the bar.  The arms are moderately tightly wound.  Arms
have many bright knots, and at low light levels there is evidence for up to
four arms, each with multiple segments.

NGC 1511:
SBd?:  Nearly edge on.  SW side of bulge is hidden by a prominent dust band,
indicating that the bulge is small.  Inner SE spiral arm has several very
bright star-forming knots.  Arms do not appear very extended (may be due to
foreshortening), but there is an extended, featureless LSB disk beyond the
arms.

NGC 1559:
SBcd (see Fig.~4e):  Moderately inclined.  Elliptical nucleus extends into
short, high surface-brightness bar.  Bar is thicker in the middle than at the
ends.  This is the only sign of a bulge.  Flocculent spiral pattern, with
evidence for at least three patchy arms.  Arms are rich in star-forming knots.
At low surface brightness levels there are a number of wispy extensions of the
arms.  In Fig.~4e, note the patches of strong dust absorption that appear as
red ($H$-band emission only) interspersed with the bright blue knots.  The
disk is much smoother in the $H$-band than the optical.

NGC 1617:
(R)SAB0/a (Type example -- see Fig.~3b):  Bright nuclear point-source.  Extends
into slightly elliptical bulge.  Weak bar.  Stubby spiral features emerge from
the bulge, but quickly fade into the overall disk emission.  Disk is slightly
asymmetric:  The major axis PA is roughly 100 degrees East of North; the disk
is slightly less extended to the north than to the south.  Disk is very smooth,
with no evidence of star-forming knots.  Disk terminates abruptly at an outer
ring.

NGC 1637:
SBb (pec):  Fairly weak bar, extending from the major axis of an elliptical
bulge.  Bar is oriented close to E-W.  Disk is highly asymmetric; much more
extended to the NE than the SW.  West arm turns to the south, and wraps 270
degrees before fading away.  East arm turns to the north, and fads after only
180 degrees.  Arms fairly well-defined, with a much higher surface brightness
on the east side.

NGC 1703:
SABb:  System close to face-on.  Symmetric nuclear source embedded in a
circular bulge.  Two-armed grand-design spiral pattern with arms emerging from
the bulge.  Arms are smooth and well-defined for $\sim$180 degrees, then become
diffuse and patchy.  One arm can be traced through a full 360 degrees.  The
other fades out after $\sim$270 degrees.

NGC 1792:
Sbc:  Centrally condensed, elliptical bulge, with no evidence for a bar.
Flocculent spiral pattern with at least five visible arms.  Arms are very
prominent, and full of star-forming knots.  There is an asymmetric feature to
the SE at low surface brightness:  One of the arms does not curve in the same
sense as the others, but instead extends practically straight along the major
axis of the disk.

NGC 1808:
SBa:  Very bright nuclear point-source embedded in a short, faint bar that is
further embedded in a boxy bulge, with a short, faint bar.  At high surface
brightness, bulge has a ``lemon'' morphology.  At lower surface brightness, the
bulge becomes strongly boxy.  Spiral arms emerge from the east and west corners
of the bulge.  The east arm has a few bright knots near the bulge, and a
diffuse patch of high surface brightness further out.  The west arm forms a
very sharp, well defined ridge, and does not appear to have any knots.  The
arms are both embedded in a smooth disk.

NGC 1832:
SBab:  Bright nuclear point-source, embedded in an elliptical bulge/stubby bar.
Bright ansae at the ends of the bar.  Bulge is embedded in a lens.  Two loosely
wrapped arms emerge from the ends of the bar.  West arm is very diffuse, and
broad.  East arm is better defined.  Both arms fade from view after winding
less than 180 degrees.  Arms are patchy, but show no signs of star-forming
knots.  The optical types for this galaxy are SB(r)bc (\markcite{rc3}RC3),
SBb(r)I (\markcite{cag}CAG), and SAB(r)c (OSU$B$).  The substantially earlier
type in the near-IR is due to the enhanced prominence of the bulge in the
$H$-band compared to the optical.

NGC 1964:
(R)Sab:  Foreground star $\sim$4$''$ from galaxy nucleus.  System is fairly
inclined.  Bright nuclear point-source embedded in a thin, HSB bar within a
bright elliptical bulge.  At lower surface brightness, the bar and foreground
star make the bulge isophotes appear lemon-shaped.  Faint, smooth spiral arms
emerge from the minor-axis ends of the bulge, and wrap for only $\sim$90
degrees before fading into a general inner disk.  The inner disk has a number
of knots and patches.  There is a faint outer disk, with a patchy outer ring
structure.

NGC 2090:
Sb (Type example -- see Fig.~3e):  Centrally condensed bulge, with no evidence
of a bar.  Two-armed, grand design spiral.  Arms are well defined, and wrap
more than 360 degrees before fading.  Outer arms become patchy.  Only a few
knots in the inner arms.

NGC 2139:
SBdm:  No bulge.  Asymmetric bar, oriented E-W (west end is brighter than east
end).  Diffuse, patchy LSB spiral features appear very disorganized.  Direction
of arm winding appears to change from the inner to the outer disk.  Two armed
pattern appears in outer disk.  Disk (and outer arms) has numerous star-forming
knots.

NGC 2196:
S0/a:  Bright nuclear point-source embedded in bright elliptical bulge.  A very
low surface brightness disk surrounds bulge.  The disk is smooth, with only a
hint of some patchyness in the NW.

NGC 2207:
SB0/a:  Interacting pair with IC 2163.  NGC 2207 is the western system.  Very
bright, elliptical bulge, with short, high surface-brightness arms appearing
roughly perpendicular to the inner major axis.  Very extensive low surface
brightness arm/ring features probably caused by tidal interaction with
companion to the east.  LSB arms have many star-forming knots.  In the SBc
interpretation, the stubby arms form ansae at the end of an LSB bar.

NGC 2280:
SBbc:  Small bright nucleus embedded in a short, faint bar.  Spiral arms emerge
from the minor axis ends of the bulge.  Within the inner disk the arms are HSB
and narrow.  They abruptly become LSB and diffuse at the edge of the inner
disk.  Both arms are smooth.  The east arm can be traced through $\sim$540
degrees.  The west arm fades after $\sim$360 degrees.  The outer arms are
knottier than the inner arms.

NGC 2442:
SBbc:  Bright nuclear point-source embedded in a small, very elliptical bulge.
Outer bulge isophotes become boxy (actually rhomboid).  Spiral arms emerge from
the obtuse corners of the rhombus.  Spiral arms are rich in star-forming knots.
Dust lanes still visible in H-band.  The arms are asymmetric.  The north arm
emerges from the east corner of the bulge, is nearly straight for some
distance, and then bifurcates, and turns sharply to the west (the bending angle
is $\sim$120 degrees).  The south arm is straight, with a narrow central ridge
line for even further, and then bends sharply to the east (the bending angle is
$\sim$150 degrees), and becomes quite diffuse and broad.  It appears to bend
entirely back on itself at low intensity level.

NGC 2559:
SBb:  Small elliptical bulge, with variable position angle:  central isophotes
are oriented roughly N-S.  At lower intensities, the PA rotates to NE-SW.
Prominent bar at same orientation as outer bulge isophotes.  Two armed spiral
pattern, with the arm emerging from the ends of the bar.  Both arms are
peculiar, with abrupt kinks in several places.  East arm has a handful of
bright star-forming knots.  West arm appears smooth.  The optical types for
this galaxy are SB(s)bc: pec (\markcite{rc3}RC3) and SABcd? (OSU$B$).  The
substantially earlier type in the near-IR is due to the heavy dust obscuration
in the central part of the galaxy which conceals much of the bulge visible in
the $H$-band.

NGC 2566:
(R)SBa:  Nuclear point-source embedded in a large slightly elliptical bulge.
Bulge is threaded by a long, thin, bright bar.  Underlying LSB disk, with an
outer ring/tightly wound spiral arms.  The ring is irregular and lumpy, but has
no obvious bright knots of current star formation.  The disk PA is offset
$\sim$30 degrees away from that defined by the bar.  The outer disk isophotes
appear offset from the nucleus of the galaxy.  However, sky irregularities make
the disk hard to characterize.

NGC 2775:
S0/a:  Very bright nuclear point-source embedded in large slightly elliptical
bulge.  Disk extends from bulge with nearly the same isophotal shape.  There is
evidence for two very faint, symmetric spiral dust absorption bands in the
outer disk.

A0908-08:
Sc?:  System nearly edge-on.  Bright, very flattened nucleus, but no sign of a
bar.  The inner disk is patchy, with a number of bright knots.  It is clearly
crossed by a dust lane on the east side.  There is a clear break between the
inner and outer disks, and the major axis PA of the disk changes at this
surface brightness discontinuity.  The high inclination make any spiral pattern
hard to detect.

NGC 2964:
SABb (pec):  Slightly elliptical bulge, with possible evidence of a weak bar
oriented close to bulge minor axis.  Four-armed spiral pattern, with two arms
emerging from ansae at the bar ends, and two others emerging near the center of
the bar.  Arms are irregular, with the outer disk having roughly rectangular
isophotes.  The two arms coming off the bar ends (the north and south arms) are
higher surface brightness than the other two.  There is some evidence for
bright knots in all but the west arm.

NGC 3059:
SBcd:  Small, very flattened bulge, threaded by a short, thin, HSB bar.  Two
loosely wrapped, smooth, LSB arms emerge from the ends of the bar.  NW arms is
much less curved than SE arm.  Outer disk isophotes are roughly circular.

NGC 3166:
SB0:  Very bright, very elliptical nuclear region, extending into similarly
flattened bulge with the major axis oriented nearly E-W.  LSB bar oriented
along bulge minor axis.  At fainter isophotes, the bulge engulfs this bar, and
becomes nearly circular.  At still fainter isophotes, the bulge recovers its
original E-W major axis, and becomes quite boxy.  Beyond this, there is an
LSB disk, with no evidence for spiral features or star forming knots.

NGC 3169:
S0/a:  Very bright, flattened nucleus, with evidence for a weak bar, embedded
in a luminous elliptical bulge.  Outer bulge isophotes appear slightly boxy.
There is a LSB disk with patchy, flocculent spiral features.  No clear arm
pattern.  Although the disk is not HSB, it is large --- perhaps overfilling the
field of view.  The optical types for this galaxy are SA(s)a pec
(\markcite{rc3}RC3), Sb(r)I-II tides (\markcite{cag}CAG), and Sb (OSU$B$).  The
substantially earlier type in the near-IR is due to the enhanced prominence of
the bulge in the $H$-band compared to the optical, and the lack of any visible
spiral structure in the $H$-band.

NGC 3223:
Sa:  Bright nuclear point-source, embedded in large, elliptical bulge.  Faint,
symmetric, tightly wrapped two-armed spiral pattern emerges from bulge.  Arms
have a few bright knots in them.

NGC 3227:
SBab:  Interacting pair with NGC 3226, with overlapping disk isophotes.  NGC
3227 is the system to the SE.  Bright nuclear points source, embedded in a
bulge with a clear X-distortion ({\it viz.}~\markcite{xes}Whitmore \& Bell
1988) in the inner isophotes.  The outer bulge isophotes are peanut-shaped.  A
bar threads the bulge, along its major axis.  The PA of the bulge and bar is in
the direction of the SE system.  There are a number of star-forming knots on
the SE side of the bar.  Two broad, smooth, diffuse spiral arms emerge from the
ends of the bar, and can be traced for $\sim$180 degrees.

NGC 3261:
SB(r)a:  Bright, centrally condensed nucleus, threaded by a low
surface-brightness bar.  Ends of the bar intersect two (even lower
surface-brightness) arms.  The arms do not commence at the ends of the bar, but
instead form a nearly complete inner pseudoring.  Beyond the bar, the arms
break away from the ring, and can be traced for more than 180 degrees.  They
are diffuse, and have obvious kinks in several places.  The east arm has a well
defined ridge line for $\sim$120 degrees past the end of the bar, and then
turns sharply and becomes very diffuse.  The west arm is never so well defined,
but can be traced significantly further before it fades away.

NGC 3275:
SB(r)ab:  Very bright nucleus embedded in an elliptical bulge with a short bar.
The bulge is embedded in a lens that extends to the ends of the bar.  Faint
spiral arms emerge from bar ends.  Arms have no evidence for knots.  At high
surface brightness, the major disk of the galaxy is aligned with the bar (PA is
$\sim$135 degrees), but at lower surface brightness, the PA is closer to 0
degrees.

NGC 3319:
SBd:  No bulge.  Nuclear point-source embedded in thin HSB bar.  The
point-source is SW of the bar center.  There are a number of other lumps along
the bar.  Two-armed, grand-design spiral pattern emanates from the ends of the
bar, but arms are LSB.  Arms can be traced through $\sim$240 degrees before
fading.  The knots are mainly in the outer parts of the arms:  The inner disk
is quite smooth.  The south arm has many more bright knots than does the north
arm.

NGC 3338:
SAB(r)b:  Nuclear point-source embedded in small elliptical bulge.  There is a
clear inner ring, with the spiral arms coming off the ring.  There is a
two-armed spiral pattern, but both arms appear to originate from the same spot
on the ring.  The arms have different inner pitch angles, and appear symmetric
at large radius.  Although there are a few knots in disk, the are not
associated with the spiral arms.  The arms are smooth.  They are narrow and
well defined at small radii, and become broad and diffuse after winding for
$\sim$180 degrees.  At faint levels, the arms can be traced for another
$\sim$180 degrees before fading into the sky.

NGC 3423:
Sd (Type example -- see Fig.~3i):  Very close to face-on.  LSB, slightly
elliptical bulge embedded in very LSB disk.  There are traces of spiral
structure in the disk, but no large-scale arms.  There are also numerous
star-forming knots distributed about the disk that do not lie on any of the
spiral segments.

NGC 3504:
(R)SB(r)a:  Bright nuclear point-source in a small slightly elliptical bulge.
Bulge is threaded by a thin, well defined bar.  Ansae at the ends of the bar
form the base of the spiral arms.  The arms wrap into an inner ring, and then
peel off to form a broken outer pseudo-ring.  There are a few knots in the
ansae, but little other evidence of ongoing star-formation.

NGC 3507:
SBb:  Nuclear point-source embedded in elliptical bulge.  The bulge is threaded
by a prominent bar.  Grand design two-armed spiral pattern, with the arms
emerging at right angles to the ends of the bar.  The arms are well-defined,
and generally smooth, although there are some knots in the inner arms.  The arm
emerging from the SE end of the bar is more tightly wrapped than that emerging
from the NW end.  The surface brightness of both arms drops abruptly after
$\sim$180 degrees.  The arms can be traced through another $\sim$180 degrees
before they fade into the sky.

NGC 3511:
SABbc:  Fairly weak bulge with evidence for a double nucleus.  Bulge is
elliptical, but the galaxy is at high inclination, so this may be a projection
effect.  Two-armed grand-design spiral pattern emerges from the bulge.  Arms
are very diffuse, thick, and patchy.  There are knots, but none are very
prominent.  At low surface-brightness there are short bits of other arms.

NGC 3513:
SBc:  Flattened nuclear source embedded in a thin, HSB bar.  Broad ansae at the
ends of the bar have a PA slightly skewed from that of the bar.  The spiral
arms emerge from the ansae.  The arms are broad, and patchy, but well defined.
They contain several obvious knots of star formation.  The arms can be followed
through $\sim$180 degrees.  They are not quite symmetric, with the south arm
being more open than the north arm, and the north arm showing signs of more
vigorous star formation.

NGC 3583:
SBc:  Bright nuclear source, embedded in an elliptical bulge.  Bulge is
threaded by a prominent bar with a PA roughly half-way between that of the
bulge major and minor axes.  Spiral structure commences at the radius of the
bar ends, but is quite peculiar.  One arm appears to originate near the NW
major axis end of the bulge.  It is quite diffuse here.  As this arm crosses
the PA of the bar, it becomes better defined, and brighter.  It then continues
to the SE, but rapidly becomes diffuse, and fades away.  A second arm appears
to commence from the SW end of the bar, and wraps very tightly around the south
side of the bulge.  A very low surface-brightness spur of this arm extends to
the east, making the system a 3-armed spiral.  When the second arm crosses the
PA of the bar on the east side, another arm commences, and runs parallel to the
second arm to the NW (that is, the second arm has a double structure once it
passes the SW end of the bar).  At low surface brightness, the arms become very
diffuse.

NGC 3593:
S0/a:  System close to edge-on.  Bright central point-source embedded in a
flattened luminous bulge.  Disk extends from the bulge with same PA.  There is
patchy extinction in the inner disk.  No evidence of any spiral pattern in the
disk.

NGC 3596:
SABb:  Nuclear point-source embedded in a slightly elliptical bulge.  Oval
inner disk aligned with bulge PA.  Two armed spiral pattern emerges from the
major-axis ends of the bulge.  North inner arm is higher surface brightness
than south arm.  Arms have ridges of higher surface brightness with lower
surface brightness extensions outward.  Both arms can be traced for 270--300
degrees.  Arm pattern on south side of the disk is complicated, with breaks
appearing in the ridge line of the arm.

NGC 3646:
SABab pec:  Nuclear point-source embedded in a small, very flattened bulge.  An
asymmetric inner disk surrounds the bulge.  The inner disk is more extended to
the NW than to the SE.  The NW side of the inner disk has a number of knots.
The spiral pattern is very odd.  One spiral arm emerges from the SE end of the
inner disk, and turns counter-clockwise.  Another emerges from the NW side of
the inner disk, and turns clockwise.  These two arms appear to merge on the SW
side of the galaxy, joining in a point.  The arm structure on the west side is
very well defined, and HSB, with many bright knots.  On the east side, the arms
are diffuse and patchy, with fewer knots.  There is an LSB outer disk on the
west side.  The referee suggests, and we agree, that this may be an example of
a galaxy with visible leading and trailing arms.

NGC 3675:
SB(r)a:  Bright nuclear point-source embedded in an elliptical bulge, threaded
by a low contrast bar.  Narrow, tightly-wound spiral arms form an inner
pseudo-ring, and can be traced for another full winding outside the ring.  Arms
are embedded in a clear interarm disk.  The outer arms are very patchy and
filamentary.

NGC 3681:
SB(r)0/a:  Nuclear point-source, threaded by short nuclear bar.  The bar causes
the bulge isophotes to be very elliptical.  Bulge is embedded in a smooth inner
disk, with a complete inner ring at its edge.  The disk surface brightness
drops abruptly outside of the ring.  The outer disk is LSB, and smooth, with no
evidence of any spiral structure.  The optical types for this galaxy are
SAB(r)bc (\markcite{rc3}RC3), SBb(r)I-II (\markcite{cag}CAG), and SAB(r)b
(OSU$B$).  The substantially earlier type in the near-IR is due to the lack of
obvious spiral structure in the $H$-band.

NGC 3684:
SABb:  Small, elliptical bulge, threaded by a prominent bar.  Two-armed grand
design spiral pattern with arms emerging from the major-axis ends of the bar.
Arms are fairly low-contrast.  There are some knots, especially on the west arm,
but the arms are fairly smooth.  Both arms can be traced through $\sim$300
degrees before the fade into the sky.

NGC 3686:
SBbc:  Bright nuclear point-source embedded in a very elliptical bulge.  Bulge
is threaded by a weak bar, with the bar PA skewed $\sim$30 degrees from the
bulge major axis.  Grand-design two-armed spiral pattern emerges from the ends
of the bar.  Inner arms are high-contrast, and have lots of filamentary
structure.  After winding $\sim$140 degrees, the arms become diffuse and LSB.
The arms can be traced through almost a full 360 degrees.  South arm bifurcates
near the bar end, with the outer part being LSB and diffuse.  Thus the system
is 3-armed at large radii and LSB levels.

NGC 3705:
SAB(r)ab (Type example -- see Fig.~3d):  Star only $\sim$3$''$ from nucleus.
Nuclear point-source embedded in an elliptical bulge.  Low contrast bar threads
bulge along the bulge minor axis.  Bar ends on a high-contrast asymmetric inner
ring.  Weak spiral pattern in the disk outside the ring.  The arms are patchy
and narrow.

NGC 3726:
SB(r)bc:  Nuclear point-source embedded in small very elliptical bulge.  Bulge
is threaded by long, prominent bar.  The bar ends at an inner ring.  The PA of
the ring is skewed by $\sim$30 degrees from the bar major axis.  Three spiral
arms emanate from the ring, with the north arm being the best-defined.  It can
be traced for $\sim$150 degrees.  The south arm is the brightest, and contains
the majority of the star-forming knots.  It is very broad at its base, and not
well defined.  It fades after only $\sim$90 degrees, but an LSB extension can
be traced for another $\sim$90 degrees.  The third (east) arm is very patchy
and diffuse.  After emerging from the east side of the ring, its inner portion
proceed in a nearly straight line to the NNW, and then bends sharply after
$\sim$90 degrees.  The bend is a $\sim$140 degree turn to the SW.  There are
several bright knots on this part of the arm.  The rest of the arm forms a very
broad LSB feature to the north and NE of the galaxy.  Because of this, the
system is very asymmetric at low surface brightness levels.

NGC 3810:
SABbc:  Nuclear point-source embedded in large, slightly elliptical bulge.  Two
armed spiral pattern emerges from the bulge, with well defined, high contrast
inner arms.  The arms become patchy, diffuse, and irregular after winding
$\sim$90 degrees.  They split into at least four outer arms.  Even the outer
arms are rich in bright knots.

NGC 3877:
Sbc?:  System close to edge-on.  Bright elliptical nuclear source embedded in a
small, less elliptical bulge.  Inclination and obvious dust extinction make the
details hard to see, but there is an open two-armed spiral pattern.  The disk
is full of bright knots.

NGC 3885:
S0/a:  Bright, very elliptical bulge, with a possible nuclear bar.  Bulge
becomes boxy at lower surface brightness.  Stumpy spiral features protrude from
the NW and SE corners of the bulge, but disappear very rapidly.  A low
surface-brightness disk extends from the bulge, and shows weak patchy features
that may be bits of arm structures.  No evidence for star formation.

NGC 3887:
SBab:  Fairly small, centrally condensed bulge, becoming very elliptical with
increasing radius.  Bulge is threaded by a bar with nearly the same PA as the
bulge major axis.  Two armed spiral pattern commences at the bar ends.  The
arms can be traced for $\sim$300 degrees before disappearing.  The arms do not
have knots, but they are quite lumpy (features much larger than typical
star-forming knots).  Beyond $\sim$180 degrees, the arms are traced mainly by
these lumps (the diffuse emission is largely lost in the sky).

NGC 3893:
SABbc:  Centrally condensed, slightly elliptical bulge.  Grand-design two-armed
spiral pattern emerges from the major-axis ends of the bulge.  The inner arms
are high contrast, rich with star-forming knots, and have obvious inner dust
lanes.  After winding for $\sim$250 degrees, they become broad, LSB outer arms
that can be traced for another $\sim$100 degrees before fading into the sky.

NGC 3938:
Sb:  System nearly face-on.  Centrally condensed nucleus in a bright bulge.
Grand-design two-armed spiral in the inner disk, with the arms bifurcating and
becoming very broad in the outer disk.  Arms are very knotty.  They can be
traced through at least a full 360 degrees before fading into the sky.

NGC 3949:
SABcd (Type example -- see Fig.~3h):  Very elliptical bulge.  Irregular,
flocculent spiral structure in disk, with many filaments, and knots of star
formation.

NGC 4027:
SBd (Type example -- see Fig.~3i):  No bulge.  Nucleus embedded in a bar.  Two
spiral arms emerge from the end of the bar.  South arm is short, winds less
than 90 degrees before fading, and has several large, bright knots of star
formation.  North arm is diffuse and broad, but winds through nearly 180
degrees before fading.  No obvious knots in outer north arm.  South arm, and
the beginning of north arm are embedded in a diffuse, patchy disk.

NGC 4030:
Sab:  Centrally condensed nucleus embedded in a large elliptical bulge.
Flocculent spiral pattern, with at least three arms.  Arms are patchy and
wispy, still showing signs of dust lanes in the H-band.  Inner disk has many
star-forming knots.  Outer disk appears quiescent.

NGC 4038:
S pec:  North system of The Antennae.  Knotty nuclear region, with one
prominent arm emerging on the SW.  Arm curves around to the north, and breaks
up into a rich array of star-forming knots on the SE side, as the arm
approaches the southern system.

NGC 4039:
S pec:  South system of The Antennae.  Nuclear point-source.  Weak arm emerges
on the SW, and quickly fades into an LSB smear.  NE side becomes rapidly
confused with the star-forming bridge from the north system.

NGC 4051:
SBb:  Bright nuclear point-source embedded in an elliptical bulge.  Bulge is
threaded by a prominent bar aligned with the bulge major axis.  System has a
complex, multi-arm spiral pattern, with two dominant arms that emerge from the
ends of the bar.  There are also two fainter arms.  One emerges from the west
end of the bulge.  The other appears to form via the bifurcation of the
southern main arm.  The arms are rich in substructure and star forming knots.

NGC 4062:
Sb:  Small, moderately elliptical bulge.  High contrast, grand-design spiral
pattern, with the arms emanating from the minor-axis ends of the bulge.  Arms
are narrow and smooth for the first 180 degrees, then become diffuse and
filamentary.  Both arms can be traced through a full 360 winding.

NGC 4100:
SBbc:  System close to edge-on.  Bright nucleus offset to the west of the
center of the outer bulge isophotes.  Bulge is threaded by a narrow,
well-defined bar.  Initially two-armed spiral pattern begins from bar ends.
Arms bifurcate after winding for $\sim$180 degrees.  Arms are fairly smooth,
with a few knots on the south side.  Arms embedded in a smooth, LSB disk.

NGC 4123:
SBb:  Nuclear point-source embedded in a very elliptical bulge.  Bulge has a
prominent X-distortion, and boxy/peanut-shaped outer isophotes.  A long, high
contrast bar threads the diagonal of the boxy bulge.  Bar is asymmetric, with
the east side being narrower, and higher surface brightness.  Two-armed spiral
pattern emerges from the ends of the bar.  The arms are very wispy and patchy.
They can be traced for $\sim$180 degrees before fading into the sky.

NGC 4136:
SB(r)a:  Small, centrally condensed bulge, threaded by a short, thick bar.  Bar
ends on a full, high contrast inner ring.  The major axis of the ring is nearly
orthogonal to that of the bar.  Two open, LSB spiral arms emerge from the major
axis ends of the ring, and can be traced through $\sim$120 degrees before
fading.  The south arm is fainter than the north arm.  Most of the bright knots
are associated with the interarm disk.  The optical types for this galaxy are
SAB(r)c (\markcite{rc3}RC3), Sc(r)I-II (\markcite{cag}CAG), and SAB(r)bc
(OSU$B$).  We classify NGC 4136 substantially earlier in the $H$-band
because the outer, open spiral arms are very faint in the near-IR, leaving the
inner arm/ring structure as the spiral feature that dominates the
classification.

NGC 4145:
SBc (Type example -- see Fig.~3g):  Foreground star or offset nuclear
point-source on SE end of a diffuse nuclear bar.  If it is a foreground star,
there is no nucleus.  Bar is embedded in a large elliptical bulge with an outer
PA skewed $\sim$45 degrees from that of the bar.  Spiral arms emerge from the
ends of the bulge, at a very acute angle.  Arms are very diffuse and
filamentary, but can be traced for $\sim$270 degrees before fading into the
sky.  Many star forming knots in the outer disk.

NGC 4151:
SB0/a:  Very bright nuclear point-source (AGN) embedded in a large, bright,
circular bulge.  Bulge is threaded by a large, thick, high contrast bar.  Weak
ansae at the bar ends.  No evidence for spiral structure.  No evidence for
star formation.

NGC 4178:
SBc?:  System nearly edge on.  No nucleus or bulge, but a long, high contrast
bar.  Two broad, low-contrast spiral arms emerge from the ends of the bar.
These fade into the ambient disk after $\sim$120 degrees.  There are three very
bright star-forming knots on the SW side of the galaxy, with many fainter ones
throughout the disk.

NGC 4212:
SABbc:  Nuclear point-source embedded in an elliptical bulge.  There is a
smooth inner disk from which two spiral arms emanate.  The west arm has higher
contrast and more knots.  It can be traced for $\sim$180 degrees before fading.
The east arm is much lower surface brightness, but can be traced $\sim$270
degrees before fading.  As the east arm turns to the north, this gives the
outer disk a strong north asymmetry.  Dust lanes are clearly visible in the
east arm.

NGC 4242:
SBm:  Faint nuclear point-source embedded in a very low surface brightness disk.
The central disk is elongated, and at considerably higher surface brightness
than the outer disk.  The outer disk is very LSB.  It has a few patches and
knots, but no sign of spiral structure.

NGC 4254:
SABc (see Fig.~4c):  Small circular bulge.  Disk has an asymmetric spiral
pattern.  The spiral arm pattern is complex.  There are three prominent arms,
and a weak fourth arm.  Two of the arms begin at the SW end of the bulge.  One
of these wraps tightly around the bulge, and the other is more open.  The third
bright arm begins at the NE end of the bulge.  The inner SW arm is just outside
the NE arm at this point.  The inner SW arm then opens, and the NE arm winds
inside it.  There is evidence for a weak arm on the west side of the galaxy.
The arms are narrow, well defined, and full of star-forming knots.  At low
surface brightness levels, the arm asymmetry makes the disk look very
lop-sided.  This is a galaxy with essentially the same morphology in the $B$-
and $H$-bands.  Note in Fig.~4c that there are no small-scale differences in
the optical and near-IR.

NGC 4293:
SBab:  System close to edge-on.  Nuclear point-source embedded in a boxy bulge
with an outer X-distortion.  A weak, short, broad bar threads the bulge along
a diagonal.  Two very open, stubby, broad spiral arms emerge from the ends of
the bar.  They can only be followed for $\sim$45 degrees before terminating
abruptly at a discontinuous decline in disk surface brightness.  The arms are
defined by a series of bright blobs.  The outer LSB disk has a PA skewed
$\sim$45 degrees from that of the inner disk/bulge.  It has traces of spiral
structure that give it an irregular isophotal shape.

NGC 4303:
SBbc:  Bright, centrally condensed circular bulge.  Bulge is threaded by a
faint bar that makes the outer bulge isophotes appear elliptical.  System has
an inner pseudoring formed by the spiral arms.  Arm pattern is complex and
irregular.  Both arms start near the bar ends, but slightly offset from them.
The south arm has a great deal of structure near its base, but becomes smooth
and diffuse after only $\sim$45 degrees.  At 90 degrees it bifurcates, with one
segment turning abruptly to the west, and then fading rapidly.  The other
segment completes the inner pseudoring on the west side.  The north arm
bifurcates immediately.  The inner segment forms the east side of the
pseudoring, and continues for a total of $\sim$240 degrees before fading.  It
has star forming knots for the first 180 degrees, and then becomes smooth and
diffuse.  The outer segments of the north arm form a wide series of knotty
filaments to the NE.  All of these fade from view after $\sim$90 degrees.

NGC 4314:
(R)SBa (see Fig.4a):  Bright, centrally condensed bulge.  Inner isophotes are
circular, but become quite elliptical at larger radii.  Bulge is boxy, and is
threaded by a long, prominent, narrow bar.  There are clear ansae at the ends
of the bar.  There is a faint ring intersecting the bar ends.  Two broad,
diffuse, LSB spiral arms also emerge from the ends of the bar.  No prominent
star-forming knots exist in the arms.  The arms can be traced for $\sim$120
degrees before fading.  Note in Fig.~4a that the only color-dependent structure
in NGC 4314 is the nuclear spiral starburst that shows up strongly in the
optical.

NGC 4388:
SB(r)a:  Elliptical nuclear region embedded in a moderately bright bulge.
System is close to edge on.  Two well-defined and tightly wrapped spiral arms
form a nearly complete ring.  The arms have a number of concentrations in them,
but these are larger and more diffuse than typical H\,II region knots.  At
fainter isophotes this bulge/ring system begins to look like a boxy bulge
embedded in a disk.  There is an LSB extension to the disk beyond the
arms/ring.

NGC 4394:
(R)SB0/a:  Bright nuclear point-source embedded in a slightly elliptical bulge.
The bulge is threaded by a prominent bar.  Ansae at the end of the bar form the
base of a tightly wrapped inner spiral arm pattern.  The arms are faint and
diffuse, although there are a number of star-forming knots visible.  The NE
arm is more prominent than the SW arm.

NGC 4414:
Sab:  Bright nuclear point-source embedded in a small elliptical bulge.  Inner
disk has a different PA than the bulge.  Spiral features appear in the inner
disk, and are brightest here.  The disk is very knotty, with evidence for many
bits of arms.  Spiral pattern is flocculent.

NGC 4448:
SB(r)a:  Very elliptical bulge, threaded by a prominent bar that is nearly
perpendicular to the bulge major axis.  Two armed spiral pattern, with arms
beginning at the ends of the bar.  Arms are very tightly wrapped, forming a
compete inner pseudoring.  System is close to edge on, and shows evidence for
a dust lane along the north side (galaxy major axis is very close to E-W).  Both
arms bifurcate after only $\sim$30 degrees, with the inner, HSB portions
forming the pseudoring.  The outer arms are much lower surface brightness, but
can still be traced for $\sim$180 degrees before fading.

NGC 4450:
SBab:  Nuclear point-source embedded in a luminous elliptical bulge.  The bulge
is threaded by a wide high-contrast bar, with a PA $\sim$20 degrees offset from
that of the bulge.  Two tightly wrapped, narrow, low-contrast arms emerge from
the ends of the bar, and can be traced through several turns before fading into
the ambient disk.  The PA of the outer disk isophotes is aligned with that of
the bulge and not that of the bar.

NGC 4457:
(R)SB0/a:  Bright, very elliptical nuclear source embedded in a nearly circular
bulge.  Bulge is threaded by a weak bar.  Bulge/bar system is embedded in a
fairly low surface brightness disk that has a clear outer ring.

NGC 4487:
SBd:  Point-source nucleus offset from the center of the inner isophotes.
Inner isophotes form a small bar/elliptical bulge.  Disk has evidence for faint
diffuse spiral features, but clear arms cannot be traced.

NGC 4490:
SBm:  System close to edge-on.  Center of system is knotty.  No nucleus, or
bulge.  Open, diffuse spiral arms emerge from central area, but quickly fade
into the ambient disk.  disk is very knotty, with obvious dust lanes and
patches.  Outer disk isophotes are irregular, with abrupt changes in PA.
Evidence for very LSB extensions of the disk to the NW and SE.

NGC 4496A:
SBd? pec:  Interacting with NGC 4496.  Faint nucleus within a thin, very
prominent bar.  There is a hint of spiral structure to the NW (away from the
brighter galaxy),  but otherwise the disk is very patchy and irregular.  There
appear to be many luminous knots/companion galaxies.  Most of these are
associated with the NW system, but they exist throughout the region.

NGC 4504:
SBcd:  Point-source nucleus.  Highly elliptical bulge, but this may be an
inclination effect.  Very low surface-brightness arms extend from the ends of
the bulge, with the PA being consistent with the bulge being weakly barred.  No
evidence of star-forming knots.  Entire galaxy is very LSB.

NGC 4527:
SABab:  Small, bright elliptical bulge.  Outer bulge isophotes are irregular,
possibly due to extinction.  Prominent grand-design two-armed spiral pattern,
with the arm origin slightly offset from the bulge major axis.  Arms are fairly
broad, even in the inner region, and show a great deal of patchy/filamentary
structure.  System is close to edge on, so foreshortening is a problem, but the
arms appear to continue for at least 180 degrees before fading.  The arms, and
the interarm regions are rich in star-forming knots.

NGC 4548:
SBab:  Nuclear point-source embedded in a slightly elliptical bulge.  Bulge has
a clear isophotal twist of $\sim$90 degrees from the inner to the outer
isophotes.  Bulge is threaded by a prominent bar, aligned with outer bulge PA.
There is an incomplete inner pseudo-ring at the ends of the bar, formed by the
inner parts of the spiral arms.  The arms do not begin at the bar ends, but
more than 90 degrees before that.  After the arms wrap past the bar, the SE arm
becomes very open, and quickly becomes diffuse.  The NW arm is much more
tightly wrapped, and remains prominent for much longer.  The east arm has many
more knots in its inner part than the west arm.  The outer disk is asymmetric,
with the SE side being higher surface brightness than the NW.  The arm
structure splits into several armlets in the outer disk.  Outer disk isophotes
are roughly aligned with the inner bulge isophotes (and orthogonal to the outer
bulge).

NGC 4568:
Sbc:  Interacting pair with NGC 4567.  NGC 4568 is the system to the SE.
Small, moderately bright elliptical bulge, embedded in an inner disk.  Inner
disk has structure, but no spiral arms.  Arms begin somewhat further out, and
are narrow, well defined and smooth.  There is a lot of interarm emission as
well.  There are fainter, flocculent arm structures on the north side of the
galaxy.

NGC 4571:
Scd (Type example -- see Fig.~3h):  Large, centrally condensed, circular bulge
embedded in an inner lens.  There is a faint two-armed spiral pattern emerging
from the lens.  The inner arms are fairly well defined.  The arm beginning in
the south is well defined for 150 degrees, and then becomes very faint and
diffuse.  The diffuse continuation can be followed for another $\sim$150
degrees.  The north arm is well defined for $\sim$270 degrees, but has several
abrupt kinks.  Beyond this, it becomes diffuse, and can be traced for another
$\sim$90 degrees.

NGC 4579:
SBa (see Fig.~4d):  Bright, centrally condensed, slightly elliptical bulge,
threaded by a prominent bar with a PA offset $\sim$30 degrees north of the
bulge major axis.  Bright ansae at the ends of the bar are the origins of the
spiral arms.  Arms are smooth, with no evidence of knots or substructure, and
tightly wrapped.  The arm emanating from the SW end of the bar is the more
prominent, and can be traced for $\sim$300 degrees.  The NE arm can be traced
for $\sim$200 degrees.  Note in Fig.~4d that the outer spiral structure in NGC
4579 is very blue.  The difference in morphology between the optical and
near-IR is driven by the very blue outer star-forming spiral arms.

NGC 4580:
S(r)b:  Small circular bulge embedded in a smooth inner disk.  The edge of the
inner disk is defined by a prominent inner ring.  The ring is formed by a pair
of spiral arms that break away after turned 180 degrees.  Once past the ring,
the arms break up into a flocculent spiral.

NGC 4593:
SBa:  Bright nuclear point-source embedded in a large elliptical/boxy bulge.  A
strong bar emerges from the NE and SW corners of the boxy bulge.  The ends of
the bar coincide with the edge of an inner disk or lens.  Two very diffuse,
smooth open spiral features emerge from the ends of the bar.  They can both be
followed for $\sim$150--180 degrees before fading.  There is evidence for a
third arm emerging from the inner disk on the south side of the galaxy.

NGC 4618:
SBm:  No bulge.  Nucleus is embedded in a diffuse bar.  The bar is not centered
on the nucleus, but rather extends further to the SW than the NE of the
nucleus.  There is one prominent spiral arm emanating from the NE end of the
bar, and continuing for at least 300 degrees before fading out.  The prominent
star-forming knots are all in the first $\sim$120 degrees.  Beyond this, the
arm becomes smooth and very diffuse.

NGC 4643:
SB0:  Bright, elliptical bulge threaded along its major axis by a long,
prominent bar.  Bulge/bar is embedded in a faint, smooth, featureless disk.

NGC 4647:
SBbc:  Small, very flat bulge embedded in a bar with the bar PA nearly
orthogonal to the bulge PA.  Symmetric two-armed spiral pattern emerges from
the ends of the bar.  Arms can be traced through a full winding before fading.
Outer arms are very fluffy.  May be effected by nearby giant elliptical.

NGC 4651:
SABa:  Bright nuclear point-source embedded in slightly elliptical bulge.
Bulge is threaded by a prominent bar that is misaligned by $\sim$30 degrees
with the bulge major axis.  A complex spiral pattern begins at the radius of
the bar ends, but the arms do not clearly begin at the ends of the bar.  There
are at least three visible, smooth, tightly wrapped and well defined arms.  Two
of them can be traced for $\sim$150 degrees.  The third is fairly short,
wrapping for only $\sim$90 degrees.  There is a clear LSB extension of the
disk, with evidence for bits of spiral structure that may or may not be
associated with the more prominent inner arms.  The optical types for this
galaxy are SA(rs)c (\markcite{rc3}RC3), Sc(r)I-II (\markcite{cag}CAG), and Sbc
(OSU$B$).  The substantially earlier type in the near-IR is due to the enhanced
prominence of the bulge in the $H$-band compared to the optical, and the very
low contrast of the spiral arms in the near-IR.

NGC 4654:
SBcd:  No real bulge.  Nucleus is threaded by a short, thin, HSB bar.  Two
prominent spiral arms emanate from the ends of the bar.  The arm emanating
from the SE end of the bar and wrapping around the north side of the disk is
the more prominent.  It is rich in star-forming knots for the first $\sim$180
degrees, and becomes smooth and diffuse thereafter.  It can be traced for
nearly a full 360 degrees.  The inner part of the NW arm is also rich in
star-forming knots, but becomes broad and diffuse after $\sim$150 degrees.  It
can only be traced $\sim$270 degrees before fading from view.

NGC 4665:
(R)SBa:  Bright round nuclear source embedded in a luminous, slightly
elliptical bulge.  Bulge is threaded by a prominent, HSB bar.  Two faint,
diffuse spiral arms emerge from the ends of the bar, and form a complete
pseudoring.  There is also clear interarm emission.  The faint outer isophotes
are clearly elliptical, at a position angle $\sim$45 degrees away from that of
the bar.

NGC 4666:
SBc?:  System close to edge-on.  Bright nucleus embedded in a thin HSB bar.
Bulge is small and flattened.  There appears to be spiral structure emanating
from the ends of the bar, but the system inclination makes the spiral pattern
hard to see.  Any spiral pattern is flocculent, rather than grand-design.  The
ends of the bar and the disk are full of star-forming knots and patches of
extinction.

NGC 4689:
Sb:  Small, slightly elliptical bulge, embedded in a slightly elliptical,
smooth inner disk.  Faint spiral features emerge from this disk at the point
that its surface brightness drops rapidly.  The arms are very low-contrast, but
clearly defined as ridges of changing surface brightness in a very LSB outer
disk.

NGC 4691:
SBd:  No real bulge.  A number of bright knots are strung along a prominent
bar.  The brightest of these knots is somewhat elliptical, and has a different
PA than does the bar, but is close to the bar mid-point.  Two diffuse, LSB
arms emerge from the ends of the bar, and for a nearly complete, faint outer
ring.  No evidence for knots associated with the arms.

NGC 4698:
S(r)a:  System seen close to edge-on.  Bright nuclear point-source embedded in
a large, luminous bulge.  Bulge isophotes are circular.  A smooth disk
surrounds the bulge.  There are no signs of star-formation, but there are
features in the disk that could be either very tightly wrapped spiral arms, or
a ring.

NGC 4699:
SB0:  Bright nuclear point-source embedded in elliptical bulge.  The bulge is
threaded by a short bar, oriented along the bulge major axis.  The bulge/bar
system is embedded in a smooth, featureless disk with the same PA as the bulge.
No evidence of spiral structure.

NGC 4772:
(R)SB0:  Nuclear point-source embedded in a bright, circular bulge.  Bulge is
threaded by a long, broad, LSB bar.  At the ends of the bar, two very faint,
narrow, smooth arms wrap around the galaxy to form an outer ring.

NGC 4775:
Sd:  Small, elliptical bulge.  Inner disk is irregular, with no clear spiral
pattern.  Spiral pattern in the outer disk is very faint and asymmetric.  The
dominant arm is very open, broad and diffuse.  It originates on the SE side of
the disk, and curves to the NW.  It can be traced for $\sim$180 degrees.  The
arm originating on the NW side of the disk is higher surface brightness, but
very short and stubby.  There are obvious knots in the arms/disk.

NGC 4781:
SBdm:  Small, faint bulge.  A small bar is oriented nearly E-W.  Very ratty,
patchy, open spiral arms emerge from the ends of the bar.  The disk has a PA
skewed about 30 degrees south of E-W (significantly different than that of the
bar).

NGC 4818:
SB0/a:  Bright nuclear point-source embedded in an elliptical bulge.  Bulge is
threaded by a bar.  The PA of the bar is skewed $\sim$30 degrees from the major
axis of the bulge.  (An alternate description is that there are short, very
loosely wrapped spiral arms emanating from the bulge).  At lower
surface-brightness, this structure simply appears to be a bulge with the PA of
the bar.  At still lower surface brightness levels, this inner structure is
embedded in a faint disk.  The PA of the disk is aligned with the PA of the
{\it inner} bulge, and not with the bar/outer bulge isophotes.  There is
evidence for very low-contrast, broad spiral features in the disk.

NGC 4856:
S0/a:  Centrally condensed, bright elliptical nucleus, with small variations in
PA.  Bulge is very large and bright, with no sign of a bar (galaxy is highly
inclined).  Stubby, smooth, low-contrast arms emerge from the bulge, and fade
rapidly into the overall disk emission.  Disk is symmetric, and featureless
aside from the stubby arms.

NGC 4900:
SBdm:  No real bulge.  Nucleus threaded by a short, HSB bar.  One prominent
spiral arm emerges from the SE end of the bar.  Arm is very patchy, and rich
in star-forming knots.  The arm is embedded in a patchy disk, the center of
which is clearly offset from the nucleus/bar.  The disk has evidence for
star-formation not associated with the arm.

NGC 4902:
SB(r)b:  Centrally condensed bulge.  Outer bulge isophotes are elliptical.  A
prominent bar is slightly skewed with respect to the bulge major axis.  Bright
stellar feature in/imposed on SW side of bar.  Bar ends in a pronounced inner
pseudo-ring.  After $\sim$150 degrees of winding, spiral arms break away from
the ring (the ring is complete, but the final $\sim$30 degrees on either side
is much fainter).  At least two fainter arms emerge earlier from the ring, the
more prominent of which emerges at a clear kink on the north side of the ring.
The arms are diffuse and patchy.

NGC 4930:
SB(r)ab:  Bright nuclear point-source embedded in an elliptical bulge, with a
long, prominent bar.  There is a resonance ring at the ends of the bar.  Ring
is slightly brighter at the ends of the bar.  Faint, smooth spiral features
extend from the ring, but are offset in PA by $\sim$135 degrees from the bar
ends (assuming trailing arms).

NGC 4939:
SABb:  Nuclear point-source embedded in a very elliptical bulge.  Evidence for
a weak bar that makes the outer bulge isophotes lemon-shaped.  There is a
tightly wrapped, two-armed grand-design spiral pattern, with the arms emerging
from the minor axis ends of the bulge.  The arms are narrow, smooth, and very
well defined.  They can both be traced through $\sim$540 degrees before fading
into the sky.  The arms become broader and more diffuse after the first
$\sim$360 degrees.

NGC 4941:
SABa:  Marginally resolved nucleus in an elliptical bulge with a strong
isophotal twist.  Smooth, low-contrast, bisymmetric two-armed spiral pattern
emerges from the major-axis ends of the bulge.  Arms can be traced for
$\sim$180 degrees before fading into the ambient disk.

IC 4182:
dI:  Nuclear point source embedded in a diffuse, LSB disk with no sign of
spiral structure, and some evidence for patchy star formation.

NGC 4995:
SBab:  Elliptical bulge, threaded by a weak bar with the bar PA skewed $\sim$45
degrees from that of the bulge.  Tightly wrapped spiral arms emerge from the
bar ends.  There are four arms, but it is really a doubled two-arm spiral,
with both arms doubled from the outset.  The inner arms are high contrast, with
sharp ridge lines.  There are a number of knots, especially on the north side
of the disk.  The outer disk isophotes are aligned with the bulge major axis.

NGC 5005:
SBa:  System is fairly inclined.  Bright, elliptical bulge with a flattened
nuclear source.  Bulge is threaded by a bar with a PA skewed $\sim$30 degrees
from that of the bulge.  The outer bulge isophotes are boxy, and the bar
crosses the diagonal of the bulge.  There are two ansae near the ends of the
bar.  Two very narrow spiral arm.  The inner arms have some evidence for
star-forming knots, and are quite open.  However, the outer arms tighten, and
appear to wrap several times around the system.  The outer disk shows an
occasional knot, but there is no coherent star-forming pattern associated with
the outer arms.

NGC 5054:
SABc:  Bright, round nuclear source threaded by a nuclear bar.  This causes the
outer bulge isophotes to appear elliptical.  There is an inner disk, with no
spiral structure.  Beyond this, there is a prominent, grand design three-armed
spiral pattern.  The arms are all well defined, and have numerous star-forming
knots.  The arm origin points are spaced $\sim$120 degrees apart, so there is
no evidence that this galaxy has two main arms plus an anomalous arm.  The N/NE
arm has the steepest pitch angle, but all are fairly loosely wound.  All the
arms can be followed for $\sim$180 degrees at the faintest level.  Beyond
$\sim$120 degrees, the N/NE arm is quite faint.

NGC 5078:
Sa:  System close to edge-on.  Bright elliptical bulge, clearly cut by an
absorption lane to the SW.  (The major axis is roughly NW-SE).    A feature
resembling a bar is probably the edge-on disk.  Very large B/D ratio, with no
clear evidence of structure aside from the prominent dust lane.  At low surface
brightness levels, the isophotes take on a pointy-box morphology typical of
disk systems with boxy bulges.

NGC 5085:
SABbc:  Small, bright bulge.  Outer bulge isophotes are elliptical, but no other
evidence of a bar.  Grand-design two-armed spiral, with well defined, broad
spiral arms.  Arms can be traced nearly 360 degrees before fading.  Arm that
emerges on the SW of the bulge bifurcates after winding almost 180 degrees.
Both arms have many star-forming knots, even quite far out in the disk.

NGC 5101:
(R)SB0/a:  Bright nuclear point-source embedded in a large, bright, circular
bulge.  At fainter isophotes the bulge becomes elliptical, and is threaded by a
thin, prominent bar.  There are ansae at the ends of the bar that form the base
of a ring/spiral arms.  No evidence of star formation, or structure in the
disk.

NGC 5121:
S0:  Bright nuclear point-source.  Inner bulge isophotes are circular, but
become elliptical at larger radii.  No evidence for a bar.  Very LSB disk
with no evidence for internal structure.

NGC 5161:
SBab:  Centrally condensed elliptical bulge embedded in a faint bar.  Bright
ansae are offset slightly downstream (assuming trailing arms) from the ends of
the bar.  Spiral pattern is two-armed, but the arms are patchy, diffuse, and
fragmented.  Arm features are clear for more than 180 degrees of winding.
The optical types for this galaxy are SA(s)c: (\markcite{rc3}RC3), Sc(s)I
(\markcite{cag}CAG), and Sc (OSU$B$).  The substantially earlier type in
the near-IR is due to the enhanced prominence of the bulge in the $H$-band
compared to the optical, and the low surface brightness of the outer, open arms
in the near-IR.

NGC 5247:
Sc (Type example -- see Fig.~3g):  Elliptical, centrally condensed bulge.  No
evidence of a bar.  Inner HSB disk shows the origin of the two-armed
grand-design spiral pattern.  Spiral arms are very well defined, with numerous
star forming knots along their central ridge-lines.  Pattern is slightly
asymmetric, with the north arm being more open and less extended than the south
arm.  Both arms bifurcate after wrapping for about 180 degrees.  The secondary
arms are much fainter than the primary arms beyond the bifurcation.

NGC 5248:
Sb:  Nuclear point-source embedded in a short, wide nuclear bar.  Nucleus/bar
structure is further embedded in a luminous elliptical bulge with a PA skewed
$\sim$45 degrees from that of the bar.  Two well defined, asymmetric spiral
arms appear in the inner disk.  The arms are rich in knots.  The north arm
turns through $\sim$90 degrees, and then terminates abruptly.  The surface
brightness of the south arm drops after $\sim$90 degrees also, but the south
arm can be traced beyond this point for another $\sim$90 degrees before fading
from view.  There are wispy, very LSB features in the outer disk, mainly on the 
west side.

ESO 383-G87:
dI (Type example -- see Fig.~3k):  No bulge.  No nucleus.  No spiral structure.
Diffuse LSB disk with an axial ratio of roughly 2 to 1.  A large number of
bright knots that could be star clusters, or individual bright stars.  The
system appears to have an isophotal twist at large radii.

NGC 5334:
SBd:  Small bulge embedded in a short, prominent bar.  The bulge/bar system is
surrounded by a wispy, irregular disk that has many spiral segments, but no
coherent spiral arm pattern.  The disk has a number of concentrations and knots
in it.

NGC 5371:
SBbc:  Nuclear point-source embedded in an elliptical bulge.  Bulge is threaded
by a prominent bar with a PA skewed $\sim$45 degrees from that of the bulge.
On the west side of the bar, a doubled arm emerges offset behind the end of the
bar.  The outer part of the doubled arm can be traced through $\sim$400 degrees
before it fades into the sky.  After $\sim$270 degrees it becomes smooth and
LSB.  On the east end of the bar a single arm emerges from the bar end, with no
offset.  The arm can be traced through $\sim$90 degrees before it merges with
the inner part of the doubled west arm.  The merged feature can be traced
through another $\sim$300 degrees after this.  For the last $\sim$120 degrees,
it is smooth and LSB.  The disk is full of knots, with the brightest knots in
the arms.  There are clear dust lanes especially on the southern side of the
disk.

NGC 5427:
Sc:  System is nearly face-on.  Bright nucleus embedded in a small, slightly
elliptical bulge.  Some evidence for a weak oval distortion with a PA skewed by
$\sim$45 from that of the bulge.  Grand-design two-armed spiral pattern emerges
from the bulge.  The arm that emerges from the west side of the bulge is HSB
for $\sim$250 degrees, and has many bright knots.  It then fades, but can be
traced through a full 360 winding.  The east arm has fewer knots, and is HSB
for only $\sim$150 degrees.  It can also be traced at lower surface brightness
levels through a full 360 degrees.  Both arms have spurs at $\sim$180 degrees.
There is a LSB bridge to the interacting companion to the south.

NGC 5448:
SABa:  System close to edge-on.  Nuclear point-source embedded in an elliptical
bulge.  Outer bulge isophotes are boxy, with a pronounced X-distortion at
low-levels.  An LSB bar threads the bulge along its diagonal.  Spiral arms
begin at the ends of the bar.  The bases of the arms are higher surface
brightness than the ends of the bar, and are also fairly broad.  The broad,
high-contrast portions of the arms are short.  After only $\sim$45 degrees, the
arms become narrow, and much fainer.  After $\sim$90 degrees the arms become
fainter still, and appear to bifurcate.  The outer LSB arms are very diffuse.
The inner arms appear to wrap around the disk, forming a nearly complete ring.

NGC 5483:
SBc:  Small bulge embedded in a small lens/bar.  Two smooth, well-defined
spiral arms emerge from the ends of the bar.  Arms emerge at right angles to
the bar PA, and can be traced for $\sim$180 degrees before fading.  Inner arms
have some evidence for knots, but outer arms are quite smooth.  Arms become
much wider, and more diffuse at $\sim$180 degrees.  PA of outer isophotes is
$\sim$90 degrees offset from that of bar/lens component.

NGC 5530:
Not observed due to the presence of a very bright star superposed on the galaxy
nucleus.

IC 4402:
Sc?:  System close to edge-on.  Small, bulge/nucleus, with no sign of a bar.
There appear to be two main spiral arms, but they are quite asymmetric.  The SE
side of the galaxy is much brighter than the NW side.  This may be due to the
dust absorption features that are obvious even in the $H$-band.

IC 4444:
(R)SB(r)cd:  Star just north of nucleus.  Small, faint bulge, threaded by well
defined bar.  There appears to be a complete inner ring at the radius of the
bar ends, but the star makes this difficult to determine.  There are at least
three spiral arms, all of which commence from the ends of the bar.  The SE end
of the bar has one arm emanating from it, winding to the west.  This arm
extends only 90 degrees before it fades into the ambient disk.  The NW end of
the bar has two arms emanating from it.  One is tightly wound, short, and full
of bright star forming knots.  The other is quite loosely wound, and wraps for
$\sim$180 degrees before fading.  This third arm has a region of intense star
formation roughly east of the bulge (winding angle of $\sim$120 degrees).

NGC 5643:
SB(r)a:  Bright nuclear point source, embedded in slightly elliptical bulge.
Bulge is threaded by a narrow, prominent bar.  Bar and bulge major axis are
misaligned by $\sim$30 degrees, but there is an isophotal twist in the bulge,
such that the bar and bulge isophotes align at low intensity.  Bright spiral
arms begin at the ends of the bar.  The spiral structure is flocculent and
asymmetric, but features can be traced out for several windings.  The arms are
rich in star-forming knots, and filamentary structure.

NGC 5676:
SABbc:  Small elliptical/boxy bulge threaded by a short, low-contrast bar.  Two
armed spiral pattern emerges from the ends of the bar, at an angle of $\sim$150
degrees to the bar major axis.  The arms are knotty, and have a number of bends
in them, but are well defined, high contrast features for almost a full 360
degrees.  Beyond this, the arm on the NE side of the galaxy stops abruptly.
The arm on the SW side becomes broad and diffuse, fading into the outer disk.
There are two short arm-spurs on the SW side of the disk.  These branch out
from the arm that ends on the NE side of the galaxy.

NGC 5701:
(R)SB0:  Bright centrally condensed nucleus embedded in a large, bright,
slightly elliptical bulge.  Bulge is threaded by a faint, diffuse bar, oriented
parallel to the bulge major axis.  Bulge/bar system is embedded in a faint
disk.  The disk is largely featureless, but there are a few objects that are
either knots or companion galaxies.  The disk may have a faint outer ring.

NGC 5713:
SBdm:  System nearly face-on.  Bright flattened nucleus, embedded in a short
HSB bar.  The bar has LSB extensions that are skewed from the main bar PA.  The
disk is very asymmetric:  There are three spiral arms, all of which emanate
from the south side of the bar (bar is oriented close to E-W).  The inner disk
and inner arms are full of knots and HSB patches.  These are much more
prevalent on the south side of the bar.  The arm structure can be traced
through $\sim$270 degrees, but becomes diffuse after the first $\sim$90
degrees.  The outer disk isophotes are not centered on the nucleus, but instead
appear off-set to the south.

NGC 5850:
SB(r)0/a (Type example -- see Fig.~3b):  Elliptical nuclear source.  The outer
bulge isophotes are nearly circular.  Bulge is threaded by a prominent, and
very long bar with a PA that is skewed by 60 degrees with respect to that of
the inner bulge.  Bright ansae at the ends of the bar.  Extensions from the
ansae form a complete inner ring.  After the ring is complete, these extensions
continue, and become faint spiral features at larger radii.  They appear to
form an incomplete outer pseudoring as well.

NGC 5921:
SB(r)bc:  Marginally resolved nucleus embedded in slightly elliptical bulge.
Bulge is threaded by a prominent bar with a PA skewed by $\sim$30 degrees from
the bulge major axis.  A sharp inner ring encloses the bar.  The ring is formed
by two spiral arm that begin at the bar ends, and nearly close together.  Just
before the ring would close, the arms become very open, broad and diffuse.  At
the north end of the bar, the arm becomes brighter as it leaves the ring.  The
outer arms are asymmetric, with the west arm being shorter and more open.

NGC 5962:
SBab:  Small, slightly elliptical bulge, threaded by a high-contrast bar.  The
bar PA is skewed $\sim$60 degrees from the bulge major axis.  Spiral arms
commence from the ends of the bar, and form an almost complete inner
pseudo-ring before opening.  The arms are fairly low-contrast and patchy.  Past
the ring, they can only be followed for $\sim$90 degrees before they fade into
the outer disk.  The outer disk has the same PA as the bulge.

NGC 6215:
SABbc:  Small bulge is elliptical at high surface brightness, becoming circular
at lower surface brightness levels.  No other evidence of a bar.  Diffuse
two-armed spiral pattern emerges along the minor axis of the bulge.  The arms
are asymmetric, with the arm emerging on the south side of the bulge being the
much higher surface-brightness arm.  There are some clear knots in the south
arm.  Both arms are doubled.

NGC 6221:
SBb:  Nuclear point-source embedded in slightly elliptical, moderately bright
bulge.  Bulge is threaded by a bar with an absorption band through it.  The bar
is aligned with the bulge minor axis.  Two very open arms emerge from the ends
of the bar.  The arms have short HSB segments that are narrow, and rich in
knots.  The north arm has a long, broad LSB continuation.  The south arm seems
to stop abruptly, but may continue at very LSB levels.  The optical types for
this galaxy are SB(s)c (\markcite{rc3}RC3), Sbc(s)II-III (\markcite{cag}CAG),
and SABcd (OSU$B$).  The substantially earlier type in the near-IR is due to
the heavy and irregular dust obscuration in the central part of the galaxy
which conceals much of the bulge visible in the $H$-band.

ESO 138-G10:
S0:  Heavy foreground contamination.  Centrally condensed nucleus, surrounded
by diffuse, featureless disk.  There is one concentration of knots to the east,
but this may be a dI companion.  The optical types for this galaxy are SA(s)dm
(\markcite{rc3}RC3) and Sdm (OSU$B$).  The radically earlier type in the
near-IR is due to the smooth, featureless appearance of the disk in the
$H$-band.

NGC 6300:
SB(r)a:  Slightly elliptical nucleus, and bulge.  Strong bar is skewed to bulge
major axis by $\sim$45 degrees.  Bar intersects a ring/tight arms.  Bright
ansae at the intersections.  Ring/arms are patchy, but fairly well defined.
Ring/arms clearly have many knots, despite obvious contamination by foreground
stars.  The optical types for this galaxy are SB(rs)b (\markcite{rc3}RC3),
SBb(s)II pec (\markcite{cag}CAG), and SAB(r)bc (OSU$B$).  The substantially
earlier type in the near-IR is due to the enhanced prominence of the bulge in
the $H$-band compared to the optical.

NGC 6384:
SBbc:  Very elliptical bulge threaded by a short, prominent bar.  The bulge
isophotes show a clear (if small) isophotal twist such that the outer bulge
isophotes are aligned with the bar major axis.  The bulge/bar system is
embedded in a lens, with a flocculent spiral pattern emerging from the lens.  A
large number of armlets are visible.  The armlets are typically smooth, with no
strong evidence of knots or dust lanes.

NGC 6753:
Sa:  Bright nuclear point-source, embedded in slightly elliptical bulge.  No
sign of a bar.  Bulge is embedded in a lens.  Beyond the lens, there are a
number of faint, tightly wrapped spiral arms embedded in a LSB disk.  Very few
star-forming knots visible.

NGC 6782:
(R)SB0/a:  Bright, centrally condensed bulge.  Nuclear region of bulge is
elliptical (major axis oriented NE-SW), but embedded in a circular bulge.
Prominent bar oriented nearly N-S.  Weak ansae at the ends of the bar signal a
faint inner disk.  Inner disk is quite elliptical, with the same major axis PA
as the bar.  Evidence for a much fainter outer disk that is nearly circular.
Arc-like features to the south appear to be due to scattered light from a
bright star.

ESO 142-G19:
Sa?:  Nearly edge-on.  Bright nuclear point-source embedded in a large,
luminous bulge.  Bulge is cut on the west side by a prominent absorption lane.
A thin, well defined disk extends from the bulge major axis.  The absorption
band can be traced out into this disk.  There appear to be a few knots of star
formation in the inner disk.

NGC 6902:
SB(r)a:  Bright, circular, centrally condensed bulge.  Bright bar oriented
NW-SE.  Two-armed spiral pattern, with arms emerging from ends of the bar, and
wrapping tightly counterclockwise.  Arms fade drastically after 180 degrees,
but continue for more than a full turn at low surface brightness.  Inner arms
appear to form an inner pseudo-ring.

NGC 6907:
SBbc:  Two-armed, grand-design spiral.  May be influenced by companion that is
superposed on NE arm.  Bulge is elliptical, and skewed with respect to the arms
(inner ``arms'' may be a bar):  The bulge major-axis is offset nearly 45
degrees from the base of the arms.  Inner arms are well defined, and have a
number of bright knots in them.  Knots apparent to much larger distances along
NE arm, possibly due to interaction with the projected companion.  At low light
levels, the arms wrap entirely around the galaxy to form an outer
pseudo-ring.

IC 5052:
Sd:  System edge-on.  No evidence of a bulge, or of spiral structure.  Despite
the edge-on orientation, there is no sign of any prominent dust lane.  The
system is oriented NW-SE; the NW side has a much higher surface brightness than
the SE side.  There are obvious knots throughout the disk.  Again, these are
more prevalent on the NW side.

NGC 7083:
Sb:  Bright, centrally condensed bulge.  Inner bulge is circular, but becomes
quite elliptical at larger radii.  Two-armed grand design spiral pattern with
the arms beginning at the bulge minor axis.  The arms each have a single
dominant ridge line, and many fainter ones.  The arms both bifurcates
immediately, with the brighter parts wrapping more tightly.  These HSB
portions wrap for $\sim$120 degrees.  The outer, LSB arms can be followed
for at least 270 degrees.  They are patchy, and have multiple armlets and
filaments.  There is obvious star formation very far out.

NGC 7184:
SB(r)a:  Bright point-like nucleus.  Elliptical bulge, with major axis aligned
with the disk major axis.  A bar appears oriented perpendicular to the bulge,
although the high inclination makes this uncertain.  As the bulge isophotes
become more circular, they connect with a bright ring, formed from tightly
wrapped spiral arms.  Arms are fairly smooth.  Isophotes indicate that the arms
begin on the SE and NW sides of the bulge, and sweep around to the NE and SW,
respectively.  More arm features, at lower surface brightness, continue beyond
the ring.

NGC 7205:
SABb (Type example -- see Fig.~3e):  Evidence for weak a bar.  Grand-design
spiral pattern that makes at least a full winding.  Arm that originates from
the SW side of the bulge is more clearly defined.  Both arms show evidence of
bifurcation/rattyness after about 180 degrees.

NGC 7213:
S0 (Type example -- see Fig.~3a):  Very bright nuclear point-source embedded in
a circular bulge.  Disk fills the frame, but is very faint, and featureless.

NGC 7217:
S0/a (Type example -- see Fig.~3b; see Fig.~4b):  Bright nucleus embedded in
very prominent, slightly elliptical bulge.  No evidence of a bar.  Faint disk
extends beyond bulge, with just a hint of wispy spiral structure in the disk.
Note in Fig.~4b that there is clear outer arm structure visible in the
$B$-band.  This indicates the arms are regions of active star formation, rather
than being dust arms.

NGC 7314:
No $JHK$ data

NGC 7410:
No $JHK$ data

NGC 7412:
SABc (Type example -- see Fig.~3g):  Small, centrally condensed, slightly
elliptical bulge.  Bulge is embedded in a smooth inner disk.  A grand-design
two-armed spiral pattern emerges from the inner disk, with the arm origins
aligned with the bulge major axis.  Both arms bifurcate.  The north arm has a
great deal of structure in its inner part.  There is a faint galaxy that may be
a companion superposed on the very inner part of the north arm.  The arm
bifurcates after $\sim$60 degrees, with the higher surface brightness part
wrapping tightly.  The HSB part can be traced for another 90 degrees.  The LSB
part is very loosely wrapped, but can still be traced for another 90 degrees.
The south arm has much less inner structure.  It also bifurcates, but after
only $\sim$30 degrees.  The inner part is also the HSB part, but here the HSB
part can be traced for $\sim$150 degrees before fading.  The LSB part fades
after only $\sim$60 degrees.  The LSB south arm shows clear evidence for
current star formation.

NGC 7418:
SBbc (see Fig.~4f):  Bright point-source nucleus embedded in a strong bar.
Complex, multi-arm spiral with arms beginning around bar ends.  At least four
arms are well enough defined to call such.  Inner parts of arms have many
bright knots.  Outer arms are fairly smooth.  Note in Fig.~4f that the spiral
arms are very blue, but the bar is a mix of green ($R$-band) and red
($H$-band).  Optical classifications of this galaxy are unbarred or weakly
barred, while it is strongly barred at $H$.

IC 5267:
No $JHK$ data

IC 5271:
No data.

IC 5273:
No data.

NGC 7479:
(R)SBbc:  Strong, thin bar, at a PA slightly skewed from the bulge major axis
PA.  Bar is close to N--S.  Bulge is elliptical.  Arms come from ends of the
bar, but are not perpendicular to the bar.  South arm commences with a sharp
west kink at the end of the bar.  Second kink to the north, and then a smooth
curve until the arm fades after roughly 180 degree.  North arm emerges smoothly
from bar, and curves smoothly back to the south.  Also fades after $\sim$180
degrees.  Arms appear to form an off-center pseudo-ring.  South arm is more
open than the north arm (thus the bulge is to the east of the center of the
pseudo-ring).

NGC 7496:
No data.

NGC 7513:
No data.

NGC 7552:
(R)SBa:  Irregular nuclear region.  At lower intensity, bulge becomes regular
and elliptical.  Prominent, long bar aligned with bulge major axis.  Bar
surface brightness is irregular, with several knots/lumps of enhanced
brightness.  Bar appears to end in points, with faint, very open spiral arms
emerging from the bar ends.  The arms are quite smooth, with few knots or
irregularities.  At low surface brightness levels, the winding of the arms
tightens, and they wrap at least 180 degrees around the galaxy.

NGC 7582:
(R)SBb?:  Bright nuclear point-source embedded in large bulge.  System is
nearly edge-on.  The inner bulge isophotes are elliptical.  At lower intensity,
the bulge has a strong ``X'' feature.  The high inclination makes any spiral
pattern impossible to detect.  However, there is obvious structure in the disk,
with clumps of star formation, and dust obscuration visible.  No coherent dust
lanes are apparent.  At low surface brightness levels, the X structure
resembles a peanut bulge, and the disk shows a pronounced integral-sign warp.
There are very LSB spiral features extending from the disk, and wrapping nearly
entirely around the galaxy.  These features are broad and smooth, with no sign
of internal structure.

NGC 7606:
Sab (Type example -- see Fig.~3d):  No sign of a bar.  Bright nuclear point
source extends into prominent elliptical bulge (ellipticity appears due to
inclination).  Generally two-armed pattern, but with several arm-fragments.
Arms are fairly smooth, with only a few bright knots.  Possible to trace arms
nearly a full 360 degrees.

IC 5325:
SABbc (Type example -- see Fig.~3f):  Nearly face-on.  Bulge is circular at
high light-levels, but shows signs of an oval distortion at lower light-levels.
Flocculent spiral pattern, with four significant arms.  Two most prominent arms
originate at the major-axis ends of the oval distortion.  West arm is higher
surface brightness than the others.

NGC 7713:
Sdm (Type example -- see Fig.~3j):  No true bulge.  Nuclear point-source
embedded in very elliptical disk.  Disk ellipticity appears due to inclination,
rather than a bar.  The HSB part of disk is not centered on the nucleus (the
nucleus is north of the center of the inner disk).  No clear spiral pattern in
the disk, although there are many star-forming knots.  The LSB part of the disk
is not centered on the HSB disk, but is more extended to the north (in the
opposite sense of the offset between the nucleus and the HSB disk).  The LSB
disk is irregular in shape, and has a number of filamentary structures in it.

NGC 7723:
SB(r)ab:  Bright, symmetric nucleus, embedded in a boxy bulge, and threaded by
a thin HSB bar.  Bar emerges from opposite corners of the boxy bulge.  Bar
terminates abruptly, and arms emerge to form an inner pseudo-ring.  Arm
structure is very complex.  One arm emerges from the SW end of the bar, with an
opening angle of $\sim$45 degrees, and winds to the NE.  It is well defined for
$\sim$90 degrees, and then becomes fat and diffuse.  The pitch angle increases
at this point.  This arm fades into the general disk after winding for
$\sim$180 degrees.  The arm associated with the NE end of the bar does not
begin at the bar.  Instead, it begins as a diffuse feature $\sim$60 degrees NW
of the bar end, and brightens until it reaches the bar end.  It then
bifurcates.  The inner (generally higher surface brightness) portion forms the
SW part of the pseudo-ring.  This inner portion bifurcates again after winding
$\sim$120 degrees from the bar end.  Once again, the inner portion forms the
pseudo-ring. And after this second bifurcation the inner arm brightens, before
fading and diffusing into the general disk.  After both bifurcations, the outer
arms are very diffuse, LSB features that wind for $\sim$90 degrees before
fading.  There are no obvious star forming knots.

NGC 7727:
S pec (early):  Elliptical nuclear source embedded in a boxy (rhomboid) bulge.
Bulge is very large.  Four faint, broad, irregular arms emerge from the bulge.
These arms have differing pitch angles, and extents.  Some can be followed
through more than 90 degrees of winding.  Some stop very quickly.  Several
large-scale LSB features are visible.  Merger remnant?

NGC 7741:
SBcd:  No true bulge.  Inner bar resolved into two sources at high surface
brightness levels.  There are many star-forming knots, mainly in the spiral arm
features.  The arms are patchy and diffuse.  The inner arms form an inner
pseudo-ring.  Outside of this ring, the arms are even more diffuse, and have
spoke-like filamentary structure.

NGC 7814:
S0/a:  Edge-on system.  Bright nuclear point-source embedded in a large,
luminous, flattened bulge, cut by an obvious dust lane along the major axis.
The disk is very thin, and imposed on the much brighter bulge.  There is no
evidence for spiral structure, due to the inclination of the disk.  There is no
evidence for star formation.  Only the presence of the dust lane prevents a
classification of S0.

\section{Summary and Discussion}

We have presented optical ($B$-band) and NIR ($H$-band) morphological 
classifications for a statistically complete sample of $\approx$200 bright 
spiral galaxies.  Our optical classifications agree well with those of the
standard catalogs (see Fig.~2a,b,c and Table 2a,b,c).  The mean absolute 
difference between our optical classifications and those from both the 
\markcite{rc3}RC3 and the \markcite{cag}CAG is about half a T-type, with no 
evidence for any systematic differences beyond those already known to exist in 
the two standard optical catalogs.  Our $H$-band classifications average  
about 1 T-type earlier than our optical classifications and the optical
classifications from both the \markcite{rc3}RC3 and the \markcite{cag}CAG (see 
Fig.~2d,e,f and Table 2d,e,f).  There is thus a clear tendency for spiral 
galaxies to appear somewhat earlier-type in the NIR than in the optical, but
with large scatter.  Our results thus do not support the assertion of 
\markcite{bnp}Block \& Puerari (1999) that the optical and NIR morphologies of 
spiral galaxies are ``decoupled''.  On average there is quite a good 
correlation between the optical and NIR morphology of spirals, although there 
are cases that show dramatic differences.

We have given short descriptions of the $H$-band morphology of our sample.  We
have also presented $B$- and $H$-band images of a set of galaxies, selected as 
examples of their $H$-band morphological type.  In addition, we have presented 
false-color ($BRH$) images of six galaxies ranging from early- to late-type 
spirals, three of which look essentially the same in the optical and the NIR, 
and three of which look substantially different.

The OSU Survey is an important tool for studies of the properties of spiral
galaxies in the nearby Universe.  The broad wavelength coverage, and the large,
well-defined and statistically complete sample invites a wide range of studies
on the morphology, dynamics, ISM, and stellar populations of spiral galaxies.
The sample also provides a template to compare with samples of galaxies at high
redshift.  Data from the OSU Survey have already contributed to a number of 
results in studies of galaxy morphology (\markcite{tea}Terndrup et al.~1994;
\markcite{qea}Quillen et al.~1997; \markcite{bar}Eskridge et al.~2000;
\markcite{ane}Elmegreen et al.~2002; \markcite{wyt}Whyte et al.~2002), galaxy
dynamics (\markcite{qfg}Quillen, Frogel \& Gonzalez 1994; \markcite{qkd}Quillen
et al.~1995; \markcite{paq}Patsis, Athanassoula \& Quillen 1997; 
\markcite{qnf}Quillen \& Frogel 1997; \markcite{pea}Puerari et al.~2000), 
galaxy evolution (\markcite{vea}van den Bergh et al.~2002), and dust and 
extinction (\markcite{knt}Kuchinski \& Terndrup 1996; \markcite{bqp}Berlind et 
al.~1997; \markcite{kea}Kuchinski et al.~1998).

We are engaged in a number of collaborative projects using the OSU Survey data. 
It is particularly important to extend the qualitative study we have done in 
this paper to more quantitative studies of galaxy morphology.  This work will 
place our understanding of the properties of current spiral galaxies on a 
firmer physical footing, and allow for a realistic comparison between the 
galaxy populations of the low- and high-redshift Universe.  We are engaged in a 
collaboration to apply the ``Hubble-space'' analysis of \markcite{anm}Abraham 
\& Merrifield (2000) to the OSU sample (\markcite{vea}van den Bergh et
al.~2002; \markcite{wyt}Whyte et al.~2002).  As the OSU sample is complete and 
well-defined, it provides for a much better assessment of local galaxy 
properties than does the sample used by \markcite{anm}Abraham \& Merrifield 
(2000).  The application of the ``Hubble-space'' analysis to our sample allows
a better analysis of the claim (\markcite{a99}Abraham et al.~1999) that the bar
fraction of disk galaxies is a strong function of redshift (\markcite{vea}van
den Bergh et al.~2002).  A related project that we are pursuing is a 
comparative study of a number of the various quantitative measures of 
``bar-strength'' that exist in the literature (e.g., \markcite{ene}Elmegreen \& 
Elmegreen 1985; \markcite{a99}Abraham et al.~1999; \markcite{bnb}Buta \& Block 
2001; \markcite{onp}Odewahn et al.~2002).

The pixel-mapping technique of \markcite{a94}Abraham et al.~(1994) is a 
powerful tool for studying the star-formation history of nearby galaxies.  
Application of this technique to a large sample of the OSU survey will result 
in a more detailed understanding of the relationship between the morphology and
star-formation histories of galaxies along the Hubble sequence.  Galaxy 
asymmetry studies are also proving to be important means of probing the 
connection between star-formation history and dynamical evolution in disk
galaxies (\markcite{cbj}Conselice, Bershady \& Jangren 2000).  The OSU survey
provides an excellent zero-redshift benchmark for such studies.

It would be especially interesting to apply the methodology of the 
``dust-penetrated'' classification scheme of \markcite{bnp}Block \& Puerari 
(1999) to {\it both} our optical and near-IR data.  We could then carry out a
quantitative assessment of the qualitative results of the current study (that
spiral galaxies are, on average slightly earlier-type in the NIR that in the
optical).  Another related program we are engaged in is a comparison of the
optical and IR properties of anemic and normal spirals (\markcite{ees}Elmegreen
\& Elmegreen 1987; \markcite{ane}Elmegreen et al.~2002).

Galaxy evolution has become a practical observational study in the last
decade.  But the sort of detailed morphological work that is possible for
samples of hundreds of galaxies, each hundreds of resolution elements across is
simply impossible for samples of tens of thousands of galaxies, each tens of
resolution elements across.  The OSU sample provides a zero-redshift benchmark
for comparison with high-redshift galaxy samples.  We are involved in 
collaborations to apply both the ``Hubble-space'' and pixel-mapping techniques
of Abraham and collaborators (\markcite{a94}Abraham et al.~1994; 
\markcite{anm}Abraham \& Merrifield 2000), and the neural-net techniques of
Odewahn and collaborators (\markcite{oea}Odewahn et al.~1996; 
\markcite{onp}2002) to the OSU Sample.  It is an irony of modern studies of 
galaxy evolution that high-redshift galaxy samples are better defined and 
better observed than zero-redshift samples.  The availability of the OSU Sample 
helps to rectify this paradoxical situation.

The broad wavelength coverage of the OSU Survey invites extensions to other
wavelength regimes.  A number of the galaxies in our sample are part of the
BIMA/SONG sample (\markcite{sng}Regan et al.~2001).  It is a natural question 
to ask how the distribution of molecular gas in these galaxies correlates with 
the distribution of both the young stellar populations (as traced by the 
$B$-band), and the old stellar populations (as traced by the $HK$ bands).  
Extension of the Survey observations into the vacuum ultra-violet allows both 
the study of the youngest stellar populations, and provides a firmer footing 
for comparisons with the rest-frame UV observations of distant galaxies from 
the Hubble Space Telescope.  We have begun such a program for a few of the 
galaxies in the OSU Survey (\markcite{ezo}Eskridge et al.~2001), however, the 
small field of the of the WFPC2 on HST has restricted this work to the smallest 
systems in the survey.  The coming availability of the wider-field Advanced 
Camera for Surveys will allow us to extend the observations to a much larger 
fraction of the Survey galaxies.

The approaching launch of SIRTF will provide an opportunity to extend the
wavelength baseline for the OSU survey through the infrared.  It will be 
possible to obtain well-resolved images for the OSU sample with IRAC in the
range of 3 to 8 microns, and marginally resolved images, and total fluxes for
the longer wavelengths with MIPS.  This will allow for a much better 
understanding of the detailed physical processes relating the stellar 
populations and the ISM in spirals.

The study of galaxy morphology has been a crucial tool for our understanding of
the Universe for the last 80 years.  We are now entering an era when the early
qualitative, single wave-band approach to morphology is maturing into a
quantitative, multi-wavelength discipline.  Our understanding of the structure, 
evolution, and underlying physics of galaxies will advance with our ability to 
study their properties across the electromagnetic spectrum, and along the
evolutionary axis of redshift.  We believe the OSU Survey will play an 
important role in this advance.

\acknowledgments

We thank the many OSU graduate students who collected data for this project,
and wish to especially note the many nights of work that Ray Bertram and Mark
Wagner have devoted to the OSU survey.  We are grateful to Roberto Aviles for
obtaining many of the optical images of southern galaxies for us with the 0.9
meter telescope at CTIO.  We thank Bob Williams and Malcolm Smith, past and
present directors of CTIO, for the generous allotment of telescope time needed
to observe most of our southern sample.  We happily acknowledge the work of 
the builders of OSIRIS and the IFPS, especially Bruce Atwood, Tom O'Brien, and 
Paul Byard of the OSU Imaging Sciences Laboratory.  JAF thanks Leonard Searle 
for the observing opportunities provided by a Visiting Research Associateship 
at Las Campanas Observatory, where some of these data were obtained.  We are 
pleased to thank Ron Buta and Allan Sandage for several very useful discussions 
on galaxy morphology.  Our referee did an outstanding job of catching many of 
the errors and weaknesses in the original manuscript.  JAF acknowledges support 
from a PPARC Senior Visiting Research Fellowship (grant no.~GR/L00896) held in 
1996 at the University of Durham Physics Department.  The construction of the 
IFPS was supported by grants AST-8822009 and AST-9112879 from the National 
Science Foundation.  The construction of OSIRIS was supported by grant 
AST-9016112 from the National Science Foundation.  This work was supported by 
grants AST-9217716 and AST-9617006 from the National Science Foundation.

\clearpage

{
\baselineskip12pt
\def\tabrule{\noalign{\hrule}}
\def\pz{\phantom{0}}
\def\pb{\phantom{-}}
\def\pd{\phantom{.}}
\ 
 
\centerline{\bf Table 1 - Sample Classification}
\vskip0.3cm
 
\newbox\tablebox
\setbox\tablebox = \vbox {
 
\halign{\pz#\pz\hfil&\pz#\pz\hfil&\pz#\pz\hfil&\pz#\pz\hfil&\pz#\pz\hfil\cr
\tabrule
\noalign{\vskip0.1cm}
\tabrule
\noalign{\vskip0.1cm}
 
\ \ Name\ \  & RC3 Type & CAG type & OSU $B$ type & OSU $H$ type \cr
\noalign{\vskip0.1cm}
\tabrule
\noalign{\vskip0.3cm}
NGC 150 & SB(rs)b & Sbc(s)II pec & SABb & SBb \cr
NGC 157 & SAB(rs)bc & Sc(s)II-III & SBbc & Sbc \cr
NGC 210 & SAB(s)b & Sb(rs)I & (R)Sab & (R)SB0/a \cr
NGC 278 & SAB(rs)b & Sbc(s)II.2 & Sab & Sb \cr
NGC 289 & SB(rs)bc & SBbc(rs)I-II & SB(r)bc & SB(r)ab \cr
NGC 428 & SAB(s)m & Sc(s)III & SABd & SBm \cr
NGC 488 & SA(r)b & Sab(rs)I & Sa & Sa \cr
NGC 578 & SAB(rs)c & Sc(s)II & SABcd & SBc \cr
NGC 613 & SB(rs)bc & SBb(rs)II & SBc & SB(r)bc \cr
NGC 625 & SB(s)m? sp & Amorph/ImIII & Sm & Sm \cr
NGC 685 & SAB(r)c & SBc(rs)II & SBcd & SBcd \cr
NGC 779 & SAB(r)b & Sb(rs)I-II & Sab & SAB(r)b \cr
NGC 864 & SAB(rs)c & Sbc(r)II-III/SBbc(r)II-III & SABc & SBb \cr
NGC 908 & SA(s)c & Sc(s)I-II & Sb & Sc \cr
NGC 986 & (R$_1'$)SB(rs)b & SBb(rs)I-II & & \cr
NGC 988 & SB(s)cd: & & & SBcd \cr
IC 239 & SAB(rs)cd & & SABc & SBc \cr
NGC 1003 & SA(s)cd & & Sdm & Scd? (edge-on) \cr
NGC 1042 & SAB(rs)cd & Sc(rs)I-II & SABbc & SABc \cr
NGC 1058 & SA(rs)c & Sc(s)II-III & Sab & Sa \cr
NGC 1073 & SB(rs)c & SBc(rs)II & SB(r)c & SB(r)ab \cr
NGC 1084 & SA(s)c & Sc(s)II.2 & Sbc & Sb \cr
NGC 1087 & SAB(rs)c & Sc(s)III.3 & SBd & SBd \cr
NGC 1187 & SB(r)c & Sbc(s)II & SBc & SB(r)b \cr
NGC 1241 & SB(rs)b & SBbc(rs)I.2 & SAB(r)bc & SB(r)ab \cr
NGC 1255 & SAB(rs)bc & Sc(s)II & & \cr
NGC 1300 & (R$'$SB(rs)bc & SBb(s)I.2 & SBb & SBb \cr
NGC 1302 & (R)SB(r)0/a & Sa & (R)SABa & SB0 \cr
NGC 1309 & SA(s)bc: & Sc(s)II & Scd & SABb \cr
NGC 1317 & SAB(r)a & Sa(s) & SB(r)a & SB0 \cr
NGC 1350 & (R$_1'$)SB(r)ab & Sa(r) & (R)SAB(r)a & (R)SB(r)a \cr
NGC 1371 & SAB(rs)a & Sa(s) & SABab & SAB(r)a \cr
NGC 1385 & SB(s)cd & SBc & SBd & SBdm \cr
NGC 1421 & SAB(rs)bc: & Sc & Sd & Sc \cr
NGC 1425 & SA(rs)b & Sb(r)II & & \cr
NGC 1433 & (R$_1'$)SB(rs)ab & SBb(s)I-II & & \cr
NGC 1493 & SB(r)cd & SBc & SABc & SBd \cr
NGC 1511 & SAa: pec & Sc pec or Amorph & Sm? & SBd? (edge-on) \cr
NGC 1559 & SB(s)cd & SBc(s)II.2 & SBc & SBcd \cr
NGC 1617 & (R$'$SB(s)a & Sa(s) & Sa & (R)SAB0/a \cr
NGC 1637 & SAB(rs)c & SBc(s)II.3 & Sbc & SBb (pec) \cr
NGC 1703 & SB(r)b & & Sc & SABb \cr
NGC 1792 & SA(rs)bc & Sc(s)II & Scd & Sbc \cr
NGC 1808 & (R$_1'$)SAB(s)a & Sbc pec & SABb? pec & SBa \cr
\noalign{\vskip0.2cm}
\tabrule
}
}
\centerline{ \box\tablebox}

\newpage

\centerline{\bf Table 1 - cont.}
\vskip0.3cm

\newbox\tablebox
\setbox\tablebox = \vbox {

\halign{\pz#\pz\hfil&\pz#\pz\hfil&\pz#\pz\hfil&\pz#\pz\hfil&\pz#\pz\hfil\cr
\tabrule
\noalign{\vskip0.1cm}
\tabrule
\noalign{\vskip0.1cm}

\ \ Name\ \  & RC3 Type & CAG type & OSU $B$ type & OSU $H$ type \cr
\noalign{\vskip0.1cm}
\tabrule
\noalign{\vskip0.3cm}
NGC 1832 & SB(r)bc & SBb(r)I & SAB(r)c & SBab \cr
NGC 1964 & SAB(s)b & Sb(s)I-II & (R)SAB(r)b & (R)Sab \cr
NGC 2090 & SA(rs)c & Sc(s)II & SABb & Sb \cr
NGC 2139 & SAB(rs)cd & SBc & SBd & SBdm \cr
NGC 2196 & (R$'$)SA(s)a & Sab(s)I & Sa & S0/a \cr
NGC 2207 & SAB(rs)bc pec & Sc(s)I.2 & S(r)b pec & SB0/a \cr
NGC 2280 & SA(s)cd & Sc(s)I.2 & Sab & SBbc \cr
NGC 2442 & SAB(s)bc pec & SBbc(rs)II & SBc & SBbc \cr
NGC 2559 & SB(s)bc: pec & & SABcd? & SBb \cr
NGC 2566 & (R$'$)SB(rs)ab: pec & & SB(r)ab & (R)SBa \cr
NGC 2775 & SA(r)ab & Sa(r) & Sa & S0/a \cr
A0908-08 & SAB(rs)b: sp & & Sc? (edge-on) & Sc? (edge-on) \cr
NGC 2964 & SAB(r)bc: & Sc(s)II.2 & Sbc & SABb pec \cr
NGC 3059 & SB(rs)bc & SBc(s)III & SBcd & SBcd \cr
NGC 3166 & SAB(rs)0/a & Sa(s) & Sa & SB0 \cr
NGC 3169 & SA(s)a pec & Sb(r)I-II tides & Sb & S0/a \cr
NGC 3223 & SA(s)b & Sb(s)I-II & Sb & Sa \cr
NGC 3227 & SAB(s) pec & Sb(s) (tides) & SABb & SBab \cr
NGC 3261 & SB(rs)b & SBab(rs)I-II & SAB(r)b & SB(r)a \cr
NGC 3275 & SB(r)ab & SBab(r)I & SB(r)ab & SB(r)ab \cr
NGC 3319 & SB(rs)cd & SBc(s)II.4 & SBcd & SBd \cr
NGC 3338 & SA(s)c & Sbc(s)I-II & SAB(r)bc & SAB(r)b \cr
NGC 3423 & SA(s)cd & Sc(s)II.2 & Sc & Sd \cr
NGC 3504 & (R)SAB(s)ab & Sb(s)/SBb(s)I-II & (R)SBab & (R)SB(r)a \cr
NGC 3507 & SB(s)b & & SBb & SBb \cr
NGC 3511 & SA(s)c & Sc(s)II.8 & Scd & SABbc \cr
NGC 3513 & SB(rs)c & SBc(s)II.2 & SBcd & SBc \cr
NGC 3583 & SB(s)b & Sbc & SABc & SBc \cr
NGC 3593 & SA(s)0/a: & Sa pec & Sa & S0/a \cr
NGC 3596 & SAB(rs)c & Sbc(r)II.2 & S(r)bc & SABb \cr
NGC 3646 & RING & Sbc(r)II pec & (R)Sb & SABab pec \cr
NGC 3675 & SA(s)b & Sb(r)II & SABab & SB(r)a \cr
NGC 3681 & SAB(r)bc & SBb(r)I-II & SAB(r)b & SB(r)0/a \cr
NGC 3684 & SA(rs)bc & Sc(s)II & SABc & SABb \cr
NGC 3686 & SB(s)bc & SBbc(s)II & SABbc & SBbc \cr
NGC 3705 & SAB(r)ab & Sb(r)I-II & S(r)b & SAB(r)ab \cr
NGC 3726 & SAB(r)c & Sbc(rs)II & SAB(r)c & SB(r)bc \cr
NGC 3810 & SA(rs)c & Sc(s)II & Sbc & SABbc \cr
NGC 3877 & SA(s)c: & Sc(s)II & Sc & Sbc? (edge-on) \cr
NGC 3885 & SA(s)0/a & Sa & Sa? pec & S0/a \cr
NGC 3887 & SB(r)bc & SBbc(s)II-III & SABbc & SBab \cr
NGC 3893 & SAB(rs)c & Sc(s)I.2 & Sbc & SABbc \cr
NGC 3938 & SA(s)c & Sc(s)I & Sc & Sb \cr
NGC 3949 & SA(s)bc & Sc(s)II-III & SABbc & SABcd \cr
NGC 4027 & SB(s)dm & Sc(s)II.2 & SBd & SBd \cr
\noalign{\vskip0.2cm}
\tabrule
}
}
\centerline{ \box\tablebox}

\newpage

\centerline{\bf Table 1 - cont.}
\vskip0.3cm

\newbox\tablebox
\setbox\tablebox = \vbox {

\halign{\pz#\pz\hfil&\pz#\pz\hfil&\pz#\pz\hfil&\pz#\pz\hfil&\pz#\pz\hfil\cr
\tabrule
\noalign{\vskip0.1cm}
\tabrule
\noalign{\vskip0.1cm}

\ \ Name\ \  & RC3 Type & CAG type & OSU $B$ type & OSU $H$ type \cr
\noalign{\vskip0.1cm}
\tabrule
\noalign{\vskip0.3cm}
NGC 4030 & SA(s)bc & & Sb & Sab \cr
NGC 4038 & SB(s)m pec & Sc (tides) & S pec & S pec \cr
NGC 4039 & SA(s)m pec & Sc (tides) & S pec & S pec \cr
NGC 4051 & SAB(rs)bc & Sbc(r)II & SBbc & SBb \cr
NGC 4062 & SA(s)c & Sc(s)II-III & Sc & Sb \cr
NGC 4100 & (R$'$)SA(rs)bc & Sc(s)I-II & Sbc & SBbc \cr
NGC 4123 & SB(r)c & SBbc(rs)I.8 & SBbc & SBb \cr
NGC 4136 & SAB(r)c & Sc(r)I-II & SAB(r)bc & SB(r)a \cr
NGC 4145 & SAB(rs)d & Sc(r)II & SAB(r)c & SBc \cr
NGC 4151 & (R$'$)SAB(rs)ab: & Sab & SBa & SB0/a \cr
NGC 4178 & SB(rs)dm & SBc & SBcd & SBc? (edge-on) \cr
NGC 4212 & SAc: & Sc(s)II-III & Sbc & SABbc \cr
NGC 4242 & SAB(s)dm & Sd/SBd & SBdm & SBm \cr
NGC 4254 & SA(s)c & Sc(s)I.3 & Sc & SABc \cr
NGC 4293 & (R)SB(s)0/a & Sa & (R)Sab pec & SBab \cr
NGC 4303 & SAB(rs)bc & Sc(s)I.2 & SBbc & SBbc \cr
NGC 4314 & SB(rs)a & SBa(rs) pec & (R)SB(r)0/a pec & (R)SBa \cr
NGC 4388 & SA(s)b: sp & Sab & Sab? & SB(r)a \cr
NGC 4394 & (R)SB(r)b & SBb(sr)I-II & SB(r)ab & (R)SB0/a \cr
NGC 4414 & SA(rs)c? & Sc(sr)II.2 & Sbc & Sab \cr
NGC 4448 & SB(r)ab & Sa(late) & SAB(r)a & SB(r)a \cr
NGC 4450 & SA(s)ab & Sab pec & (R)Sa & SBab \cr
NGC 4457 & (R)SAB(s)0/a & RSb(s)II & (R)Sab & (R)SB0/a \cr
NGC 4487 & SAB(rs)cd & Sc(s)II.2 & SABcd & SBd \cr
NGC 4490 & SB(s)d pec & Scd/SBcd & SBd & SBm \cr
NGC 4496A & SB(rs)m & & SB(r)d & SBd? pec \cr
NGC 4504 & SA(s)cd & Sc(s)II & Scd & SBcd \cr
NGC 4527 & SAB(s)bc & Sb(s)II & Sb & SABab \cr
NGC 4548 & SB(rs)b & SBb(rs)I-II & SB(r)ab & SBab \cr
NGC 4568 & SA(rs)bc & Sc(s)II-III & Sc pec & Sbc \cr
NGC 4571 & SA(r)d & Sc(s)II-III & S(r)c & Scd \cr
NGC 4579 & SAB(rs)b & Sab(s)II & SB(r)ab & SBa \cr
NGC 4580 & SAB(rs)a pec & Sc(s)/Sa & S(r)b & S(r)b \cr
NGC 4593 & (R)SB(rs)b & SBb(rs)I-II & SB(r)ab & SBa \cr
NGC 4618 & SB(rs)m & SBbc(rs)II.2 & SBdm & SBm \cr
NGC 4643 & SB(rs)0/a & SB03/SBa & SB(r)0/a & SB0 \cr
NGC 4647 & SAB(rs)c & & SB(r)c & SBbc \cr
NGC 4651 & SA(rs)c & Sc(r)I-II & Sbc & SABa \cr
NGC 4654 & SAB(rs)cd & SBc(rs)II-III & SABbc & SBcd \cr
NGC 4665 & SB(s)0/a & & SB0/a & (R)SBa \cr
NGC 4666 & SABc: & SBcII.3 & Sbc? & SBc? (edge-on) \cr
NGC 4689 & SA(rs)bc & Sc(s)II.3 & Sbc & Sb \cr
NGC 4691 & (R)SB(s)0/a pec & R Amorph pec & (R)SBd pec & SBd \cr
NGC 4698 & SA(s)ab & Sa & Sa & S(r)a \cr
NGC 4699 & SAB(rs)b & Sab(sr) or Sa & SBa & SB0 \cr
\noalign{\vskip0.2cm}
\tabrule
}
}
\centerline{ \box\tablebox}

\newpage

\centerline{\bf Table 1 - cont.}
\vskip0.3cm

\newbox\tablebox
\setbox\tablebox = \vbox {

\halign{\pz#\pz\hfil&\pz#\pz\hfil&\pz#\pz\hfil&\pz#\pz\hfil&\pz#\pz\hfil\cr
\tabrule
\noalign{\vskip0.1cm}
\tabrule
\noalign{\vskip0.1cm}

\ \ Name\ \  & RC3 Type & CAG type & OSU $B$ type & OSU $H$ type \cr
\noalign{\vskip0.1cm}
\tabrule
\noalign{\vskip0.3cm}
NGC 4772 & SA(s)a & & Sa & (R)SB0 \cr
NGC 4775 & SA(s)d & Sc(s)III & Sd & Sd \cr
NGC 4781 & SB(rs)d & Sc(s)III & SABdm & SBdm \cr
NGC 4818 & SAB(rs)ab: & & S pec & SB0/a \cr
NGC 4856 & SB(s)0/a & S01(6)/Sa & S0/a & S0/a \cr
NGC 4900 & SB(rs)c & & SBd & SBdm \cr
NGC 4902 & SB(r)b & SBb(rs)I-II & SB(r)b & SB(r)b \cr
NGC 4930 & SB(rs)b & & SB(r)b & SB(r)ab \cr
NGC 4939 & SA(s)bc & Sbc(rs)I & SABb & SABb \cr
NGC 4941 & (R)SAB(r)ab: & Sab(s)II & SABab & SABa \cr
IC 4182 & SA(s)m & & Sm & dI \cr
NGC 4995 & SAB(rs)b & Sbc(s)II & SAB(r)ab & SBab \cr
NGC 5005 & SAB(rs)bc & Sb(s)II & Sb & SBa \cr
NGC 5054 & SA(s)bc & Sb(s)I-II & Sc & SABc \cr
NGC 5078 & SA(s)a: sp & & Sa? & Sa (edge-on) \cr
NGC 5085 & SA(s)c & Sc(r)II & Sc & SABbc \cr
NGC 5101 & (R)SB(rs)0/a & SBa & (R)SB(r)0/a & (R)SB0/a \cr
NGC 5121 & (R$'$)SA(s)a & Sa & S0/a & S0 \cr
NGC 5161 & SA(s)c: & Sc(s)I & Sc & SBab \cr
NGC 5247 & SA(s)bc & Sc(s)I-II & Sc & Sc \cr
NGC 5248 & SAB(rs)bc & Sbc(s)I-II & S(r)c & Sb \cr
ESO 383-G87 & SB(s)dm & & SBm & dI \cr
NGC 5334 & SB(rs)c: & SBc(rs)II & SBcd & SBd \cr
NGC 5371 & SAB(rs)bc & Sb(rs)I/SBb(rs)I & SABbc & SBbc \cr
NGC 5427 & SA(s)c pec & Sbc(s)I & S(r)c & Sc \cr
NGC 5448 & (R)SAB(r)a & Sa(s) & SAB(r)b & SABa \cr
NGC 5483 & SA(s)c & SBbc(s)II-III & SABc & SBc \cr
NGC 5530 & SA(rs)c & Sc(s)II.8 & & \cr
IC 4402 & SA(s)b? sp & & Scd? & Sc? (edge-on) \cr
IC 4444 & SAB(rs)bc: & Sc(s)II pec & Sd & (R)SB(r)cd \cr
NGC 5643 & SAB(rs)c & SBc(s)II-III & SAB(r)b & SB(r)a \cr
NGC 5676 & SA(rs)bc & Sc(s)II & Sbc & SABbc \cr
NGC 5701 & (R)SB(rs)0/a & (PR)SBa & (R)SB0/a & (R)SB0 \cr
NGC 5713 & SAB(rs)bc pec & Sbc(s) pec & SABd & SBdm \cr
NGC 5850 & SB(r)b & SBb(sr)I-II & SB(r)ab & SB(r)0/a \cr
NGC 5921 & SB(r)bc & SBbc(s)I-II & SB(r)bc & SB(r)bc \cr
NGC 5962 & SA(r)c & Sc(rs)II.3 & SABb & SBab \cr
NGC 6215 & SA(s)c & Sc(s)II & Scd & SABbc \cr
NGC 6221 & SB(s)c & Sbc(s)II-III & SABcd & SBb \cr
ESO 138-G10 & SA(s)dm & & Sdm & S0 \cr
NGC 6300 & SB(rs)b & SBb(s)II pec & SAB(r)bc & SB(r)a \cr
NGC 6384 & SAB(r)bc & Sb(r)I.2 & SB(r)bc & SBbc \cr
NGC 6753 & (R$'$)SA(r)b & Sb(r)I & (R)S(r)a & Sa \cr
NGC 6782 & (R)SAB(r)a & SBab(s) & (R)SAB(r)a & (R)SB0/a \cr
ESO 142-G19 & Sb: sp & & Sa? (edge-on) & Sa? (edge-on) \cr
\noalign{\vskip0.2cm}
\tabrule
}
}
\centerline{ \box\tablebox}

\newpage

\centerline{\bf Table 1 - cont.}
\vskip0.3cm

\newbox\tablebox
\setbox\tablebox = \vbox {

\halign{\pz#\pz\hfil&\pz#\pz\hfil&\pz#\pz\hfil&\pz#\pz\hfil&\pz#\pz\hfil\cr
\tabrule
\noalign{\vskip0.1cm}
\tabrule
\noalign{\vskip0.1cm}

\ \ Name\ \  & RC3 Type & CAG type & OSU $B$ type & OSU $H$ type \cr
\noalign{\vskip0.1cm}
\tabrule
\noalign{\vskip0.3cm}
NGC 6902 & SA(r)b & Sa(r) & S(r)ab & SB(r)a \cr
NGC 6907 & SB(s)bc & SBbc(s)II & SBc & SBbc \cr
IC 5052 & SBd: sp & Scd/SBcd & Sd? & Sd \cr
NGC 7083 & SA(s)bc & Sb(s)I-II & Sbc & Sb \cr
NGC 7184 & SB(r)c & Sb(r)II & S(r)a? & SB(r)a \cr
NGC 7205 & SA(s)bc & Sb(r)II.8 & Sb & SABb \cr
NGC 7213 & SA(s)a: & Sa(rs) & Sa & S0 \cr
NGC 7217 & (R)SA(r)ab & Sb(r)II-III & (R)Sa & S0/a \cr
NGC 7314 & SAB(rs)bc & Sc(s)III & & \cr
NGC 7410 & SB(s)a & SBa or Sa & & \cr
NGC 7412 & SB(s)b & Sc(s)I-II & SABc & SABc \cr
NGC 7418 & SAB(rs)cd & Sc(rs)I.8 & SABbc & SBbc \cr
IC 5267 & (R)SA(rs)0/a & Sa(r) & & \cr
IC 5271 & Sb? & Sb(rs)II & & \cr
IC 5273 & SB(rs)cd & SBc(s)II-III & & \cr
NGC 7479 & SB(s)c & SBbc(s)I-II & SBbc & (R)SBbc \cr
NGC 7496 & (R$'$:)SB(rs)bc & SBc(s)II.8 & & \cr
NGC 7513 & (R$'$)SB(s)b pec & & & \cr
NGC 7552 & (R$'$)SB(s)ab & SBb(s)I-II & (R)SBb & (R)SBa \cr
NGC 7582 & (R$_1'$)SB(s)ab & SBab(rs) & SBb? & (R)SBb? (edge-on) \cr
NGC 7606 & SA(s)b & Sb(r)I & Sab & Sab \cr
IC 5325 & SAB(rs)bc & Sc(s)II-III & Sbc & SABbc \cr
NGC 7713 & SB(r)d: & Sc(s)II-III & Sd & Sdm \cr
NGC 7723 & SB(r)b & SBb(rs)I-II & SB(r)bc & SB(r)ab \cr
NGC 7727 & SAB(s)a pec & Sa pec & SABa pec & S pec (early) \cr
NGC 7741 & SB(s)cd & SBc(s)II.2 & SBcd & SBcd \cr
NGC 7814 & SA(s)ab: sp & S(ab) & Sa? (edge-on) & S0/a (edge-on) \cr
\noalign{\vskip0.2cm}
\tabrule
}
}
\centerline{ \box\tablebox}
}

\newpage

{
\baselineskip12pt
\def\tabrule{\noalign{\hrule}}
\def\pz{\phantom{0}}
\def\pb{\phantom{-}}
\def\pd{\phantom{.}}
\ 
 
\centerline{\bf Table 2a - Comparison Between the RC3 and CAG}
\vskip0.3cm
 
\newbox\tablebox
\setbox\tablebox = \vbox {
 
\halign{\hfil\pz#\pz\hfil&\hfil\pz#\pz\hfil&\hfil\pz#\pz\hfil&\hfil\pz#\pz\hfil&
\hfil\pz#\pz\hfil&\hfil\pz#\pz\hfil\cr
\tabrule
\noalign{\vskip0.1cm}
\tabrule
\noalign{\vskip0.1cm}
 
RC3 type & Number & Mean CAG type & (RC3$-$CAG) & Disp & Median \cr
\noalign{\vskip0.1cm}
\tabrule
\noalign{\vskip0.3cm}
 0    &     10    &      1.0   &   $-$1.0 & 1.2  & 1    \cr
 1    &     17    &      1.9   &   $-$0.9 & 1.2  & 1    \cr
 2    &     15    &      2.1   &   $-$0.1 & 0.8  & 2    \cr
 3    &     29    &      3.0   &   $\pb$0.0 & 0.8  & 3    \cr
 4    &     42    &      4.2   &   $-$0.2 & 0.8  & 4    \cr
 5    &     40    &      4.7   &   $\pb$0.3 & 0.5  & 5    \cr
 6    &     14    &      5.0   &   $\pb$1.0 & 0$\pd\pz$    & 5    \cr
 7    &     $\pz$7    &      5.3   &  $\pb$1.5 & 0.5  & 5   \cr
 8    &     $\pz$3    &      5.7   &  $\pb$2.3 & 0.9  & 5  \cr
 9    &     $\pz$5    &      5.8   &  $\pb$3.2 & 2.1  & 5 \cr
\noalign{\vskip0.2cm}
\tabrule
}
}
\centerline{ \box\tablebox}

\vskip0.8cm

\centerline{\bf Table 2b - Comparison Between the RC3 and OSU $B$}
\vskip0.3cm

\newbox\tablebox
\setbox\tablebox = \vbox {

\halign{\hfil\pz#\pz\hfil&\hfil\pz#\pz\hfil&\hfil\pz#\pz\hfil&\hfil\pz#\pz\hfil&
\hfil\pz#\pz\hfil&\hfil\pz#\pz\hfil\cr
\tabrule
\noalign{\vskip0.1cm}
\tabrule
\noalign{\vskip0.1cm}

RC3 type & Number & Mean OSUB type & (RC3$-$OSUB) & Disp & Median \cr
\noalign{\vskip0.1cm}
\tabrule
\noalign{\vskip0.3cm}
 0    &     10     &     0.6   &    $-$0.6 & 0.7  & 0.5  \cr 
 1    &     19     &     2.3   &    $-$1.3 & 2.3  & 1$\pd\pz$   \cr 
 2    &     15     &     1.7   &    $\pb$0.3 & 0.8  & 1$\pd\pz$   \cr 
 3    &     32     &     2.8   &    $\pb$0.2 & 1.3  & 2$\pd\pz$   \cr 
 4    &     41     &     4.4   &    $-$0.4 & 1.2  & 4$\pd\pz$   \cr 
 5    &     41     &     4.6   &    $\pb$0.4 & 1.2  & 5$\pd\pz$   \cr 
 6    &     15     &     5.3   &    $\pb$0.7 & 1.5  & 5$\pd\pz$   \cr 
 7    &     $\pz$7     &     6.6   &    $\pb$0.4 & 1.1  & 7$\pd\pz$   \cr 
 8    &     $\pz$5     &     7.6   &    $\pb$0.4 & 1.1  & 8$\pd\pz$   \cr 
 9    &     $\pz$5     &     8.0   &    $\pb$1.0 & 1.0  & 8$\pd\pz$    \cr 
\noalign{\vskip0.2cm}
\tabrule
}
}
\centerline{ \box\tablebox}

\vskip0.8cm

\centerline{\bf Table 2c - Comparison Between the CAG and OSU $B$}
\vskip0.3cm

\newbox\tablebox
\setbox\tablebox = \vbox {

\halign{\hfil\pz#\pz\hfil&\hfil\pz#\pz\hfil&\hfil\pz#\pz\hfil&\hfil\pz#\pz\hfil&
\hfil\pz#\pz\hfil&\hfil\pz#\pz\hfil\cr
\tabrule
\noalign{\vskip0.1cm}
\tabrule
\noalign{\vskip0.1cm}

CAG type & Number & Mean OSUB type & (CAG$-$OSUB) & Disp & Median \cr
\noalign{\vskip0.1cm}
\tabrule
\noalign{\vskip0.3cm}
$\pz$0    &   $\pz$2   &    0$\pd\pz$   &  $\pb$0$\pd\pz$ & 0$\pd\pz$  & 0  \cr
$\pz$1     &    20     &     1.1   &    $-$0.1 & 0.8  &  1   \cr
$\pz$2     &    13     &     1.6   &    $\pb$0.4 & 0.8  &  1   \cr
$\pz$3     &    34     &     2.9   &    $\pb$0.1 & 1.1  &  3   \cr
$\pz$4     &    28     &     4.4   &    $-$0.4 & 1.3  &  4   \cr
$\pz$5     &    68     &     5.0   &    $\pb$0$\pd\pz$ & 1.4  &  5   \cr
$\pz$6     &    $\pz$2     &     7.0   &    $-$1.0 & 0$\pd\pz$   &  7   \cr
$\pz$7     &    $\pz$1     &     8.0   &    $-$1.0 & 0$\pd\pz$   &  8   \cr
10     &    $\pz$1     &     9.0   &    $\pb$1.0 & 0$\pd\pz$   &  9   \cr
\noalign{\vskip0.2cm}
\tabrule
}
}
\centerline{ \box\tablebox}

\newpage

\centerline{\bf Table 2d - Comparison Between the RC3 and OSU $H$}
\vskip0.3cm

\newbox\tablebox
\setbox\tablebox = \vbox {

\halign{\hfil\pz#\pz\hfil&\hfil\pz#\pz\hfil&\hfil\pz#\pz\hfil&\hfil\pz#\pz\hfil&
\hfil\pz#\pz\hfil&\hfil\pz#\pz\hfil\cr
\tabrule
\noalign{\vskip0.1cm}
\tabrule
\noalign{\vskip0.1cm}

RC3 type & Number & Mean OSUH type & (RC3$-$OSUH) & Disp & Median \cr
\noalign{\vskip0.1cm}
\tabrule
\noalign{\vskip0.3cm}
 0    &     10     &    $-$0.3   &    $\pb$0.3 & 0.2  &  0    \cr 
 1    &     18     &    $\pb$1.2   &    $-$0.2 & 2.3  &  1    \cr 
 2    &     16     &    $\pb$1.0   &    $\pb$1.0 & 0.9  &  1    \cr 
 3    &     32     &    $\pb$1.9   &    $\pb$1.1 & 1.5  &  2    \cr 
 4    &     41     &    $\pb$3.5   &    $\pb$0.5 & 1.5  &  4    \cr 
 5    &     41     &    $\pb$3.7   &    $\pb$1.3 & 1.6  &  4    \cr 
 6    &     16     &    $\pb$6.1   &    $-$0.1 & 1.2  &  6    \cr 
 7    &     $\pz$7     &    $\pb$7.1   &    $-$0.1 & 1.2  &  7    \cr 
 8    &     $\pz$5     &    $\pb$6.0   &    $\pb$2.0 & 3.9  &  7    \cr 
 9    &     $\pz$5     &    $\pb$8.8   &    $\pb$0.2 & 1.0  &  9    \cr 
\noalign{\vskip0.2cm}
\tabrule
}
}
\centerline{ \box\tablebox}

\vskip0.8cm

\centerline{\bf Table 2e - Comparison Between the CAG and OSU $H$}
\vskip0.3cm

\newbox\tablebox
\setbox\tablebox = \vbox {

\halign{\hfil\pz#\pz\hfil&\hfil\pz#\pz\hfil&\hfil\pz#\pz\hfil&\hfil\pz#\pz\hfil&
\hfil\pz#\pz\hfil&\hfil\pz#\pz\hfil\cr
\tabrule
\noalign{\vskip0.1cm}
\tabrule
\noalign{\vskip0.1cm}

CAG type & Number & Mean OSUH type & (CAG$-$OSUH) & Disp & Median \cr
\noalign{\vskip0.1cm}
\tabrule
\noalign{\vskip0.3cm}
$\pz$0     &    $\pz$2     &    $-$0.5    &   $\pb$0.5 & 0.5  & $-$0.5 \cr
$\pz$1     &    20     &    $\pb$0.2    &   $\pb$0.8 & 0.9  & $\pb$0$\pd\pz$ \cr
$\pz$2     &    13     &    $\pb$0.8    &   $\pb$1.2 & 1.0  & $\pb$1$\pd\pz$ \cr
$\pz$3     &    34     &    $\pb$1.8    &   $\pb$1.2 & 1.4  & $\pb$2$\pd\pz$ \cr
$\pz$4     &    28     &    $\pb$3.6    &   $\pb$0.4 & 1.7  & $\pb$3$\pd\pz$ \cr
$\pz$5     &    68     &    $\pb$4.7    &   $\pb$0.3 & 2.0  & $\pb$4.5 \cr
$\pz$6     &$\pz$2     &    $\pb$8.0    &   $-$2.0 & 1.0  & $\pb$8$\pd\pz$ \cr
$\pz$7     &$\pz$1   &  $\pb$9.0  &   $-$2.0 & 0$\pd\pz$   & $\pb$9$\pd\pz$ \cr
10     &  $\pz$1   &  $\pb$9.0   &   $\pb$1.0 & 0$\pd\pz$   & $\pb$9$\pd\pz$ \cr
\noalign{\vskip0.2cm}
\tabrule
}
}
\centerline{ \box\tablebox}

\vskip0.8cm

\centerline{\bf Table 2f - Comparison Between the OSU $B$ and OSU $H$}
\vskip0.3cm

\newbox\tablebox
\setbox\tablebox = \vbox {

\halign{\hfil\pz#\pz\hfil&\hfil\pz#\pz\hfil&\hfil\pz#\pz\hfil&\hfil\pz#\pz\hfil&
\hfil\pz#\pz\hfil&\hfil\pz#\pz\hfil\cr
\tabrule
\noalign{\vskip0.1cm}
\tabrule
\noalign{\vskip0.1cm}

OSUB type & Number & Mean OSUH type & (OSUB$-$OSUH) & Disp & Median \cr
\noalign{\vskip0.1cm}
\tabrule
\noalign{\vskip0.3cm}
 0     &    $\pz$7     &    $-$0.1   &    $\pb$0.1 & 0.9  &  0$\pd\pz$ \cr
 1     &    25     &    $\pb$0.2   &    $\pb$0.8 & 0.9  &  0$\pd\pz$ \cr
 2     &    22     &    $\pb$1.4   &    $\pb$0.6 & 1.0  &  1$\pd\pz$ \cr
 3     &    28     &    $\pb$2.0   &    $\pb$1.0 & 1.2  &  2$\pd\pz$ \cr
 4     &    35     &    $\pb$3.4   &    $\pb$0.6 & 1.2  &  4$\pd\pz$ \cr
 5     &    32     &    $\pb$4.2   &    $\pb$0.8 & 1.3  &  4$\pd\pz$ \cr
 6     &    17     &    $\pb$5.1   &    $\pb$0.9 & 1.4  &  5$\pd\pz$ \cr
 7     &    15     &    $\pb$7.4   &    $-$0.4 & 1.1  &  7$\pd\pz$ \cr
 8     &    $\pz$5     &    $\pb$6.2   &    $\pb$1.8 & 4.2  &  8$\pd\pz$ \cr
 9     &    $\pz$4     &    $\pb$9.0   &    $\pb$0$\pd\pz$ & 1.4  &  9.5 \cr
\noalign{\vskip0.2cm}
\tabrule
}
}
\centerline{ \box\tablebox}
}

\newpage

{
\baselineskip12pt
\def\tabrule{\noalign{\hrule}}
\def\pz{\phantom{0}}

\centerline{\bf Table 3 - H-Band Type Examples}
\vskip0.3cm
 
\newbox\tablebox
\setbox\tablebox = \vbox {
 
\halign{\pz#\pz\hfil&\pz#\pz\hfil&\pz#\pz\hfil&\pz#\pz\hfil&\hfil\pz#\pz\hfil\cr
\tabrule
\noalign{\vskip0.1cm}
\tabrule
\noalign{\vskip0.1cm}
 
OSU $H$ Type & \ \ \ \ SA & \ \ \ SAB & \ \ \ \ SB & Fig.~3 \cr
\noalign{\vskip0.1cm}
\tabrule
\noalign{\vskip0.3cm}
\ \ \ \ \ S0 & NGC 7213 & & NGC 1317 & a \cr
\ \ \ \ \ S0/a & NGC 7217 & NGC 1617 & NGC 5850 & b \cr
\ \ \ \ \ Sa & NGC 488 & NGC 1371 & NGC 1350 & c \cr
\ \ \ \ \ Sab & NGC 7606 & NGC 3705 & NGC 289 & d \cr
\ \ \ \ \ Sb & NGC 2090 & NGC 7205 & NGC 1300 & e \cr
\ \ \ \ \ Sbc & NGC 157 & IC 5325 & NGC 613 & f \cr
\ \ \ \ \ Sc & NGC 5247 & NGC 7412 & NGC 4145 & g \cr
\ \ \ \ \ Scd & NGC 4571 & NGC 3949 & NGC 685 & h \cr
\ \ \ \ \ Sd & NGC 3423 & & NGC 4027 & i \cr
\ \ \ \ \ Sdm & NGC 7713 & & NGC 1385 & j \cr
\ \ \ \ \ Sm & NGC 625 & & NGC 428 & k \cr
\ \ \ \ \ dI & ESO 383-G87 & & & k \cr
\noalign{\vskip0.2cm}
\tabrule
}
}
\centerline{ \box\tablebox}
}

\clearpage 

\figcaption{Histograms of the distribution of RC3 T-types for all spirals in 
the RC3 with $M_B \leq 12$ and ellipticity $< 0.5$ (solid histogram), and for 
the OSU Sample (dotted histogram).  The coding of RC3 T-type with Hubble stage
is shown along the bottom of the figure.}
 
\figcaption{Comparisons of morphological types from the RC3, the CAG, our 
$B$-band data, and our $H$-band data.  a) RC3 types versus CAG types, b) RC3 
types versus OSU $B$-band types, c) CAG types versus OSU $B$-band types, d) 
RC3 types versus OSU $H$-band types, e) CAG types versus OSU $H$-band types, f)
OSU $B$-band types versus OSU $H$-band types.  The size of the solid polygons
scales with the number of galaxies at that location.  The open circles are the
mean y-axis values for a given x-axis value, and the error bars are the 
dispersions about those means.  The large open triangles are the median y-axis 
values for a given x-axis value.  For statistical purposes, objects classified 
as S0 are scored as $T=-1$.}

\figcaption{$B$- and $H$-band images of galaxies selected as $H$-band type 
examples.  The $B$ images are on the left, and the $H$ on the right.  All
images are oriented N up, E to the left.  The scale bars on the $B$ images
are $1'$ long; each $BH$ image pair are at the same scale.  The OSU $H$-band
types are shown on the $H$-band images.  a) NGC 7213 -- S0 (top) and NGC 1317 
-- SB0 (bottom).  b) NGC 7217 -- S0/a (top), NGC 1617 -- SAB0/a (middle), and 
NGC 5850 -- SB0/a (bottom).  c) NGC 488 -- Sa (top), NGC 1371 -- SABa (middle), 
and NGC 1350 -- SBa (bottom).  d) NGC 7606 -- Sab (top), NGC 3705 -- SABab 
(middle), and NGC 289 -- SBab (bottom).  e) NGC 2090 -- Sb (top), NGC 7205 -- 
SABb (middle), NGC 1300 SBb (bottom).  f) NGC 157 -- Sbc (top), IC 5325 -- 
SABbc (middle), NGC 613 -- SBbc (bottom).  g) NGC 5247 -- Sc (top), NGC 7412 -- 
SABc (middle), NGC 4145 -- SBc (bottom).  h) NGC 4571 -- Scd (top), NGC 3949 -- 
SABcd (middle), NGC 685 -- SBcd (bottom).  i) NGC 3423 -- Sd (top), NGC 4027 -- 
SBd (bottom).  j) NGC 7713 -- Sdm (top), NGC 1385 -- SBdm (bottom).  k) NGC 625 
-- Sm (top), NGC 428 -- SBm (middle), ESO 383-G87 -- dI (bottom).}

\figcaption{Color-composite images of representative galaxies.  Blue shows the
$B$-band, green the $R$-band, and red the $H$-band.  a) NGC 4314, b) NGC 7217,
c) NGC 4254, d) NGC 4579, e) NGC 1559, f) NGC 7418.}


\clearpage

\begin{references}
\reference{a99}Abraham, R.G., Merrifield, M.R., Ellis, R.S., Tanvir, N.R. \&
Brinchmann, J. 1999, \mnras, 308, 569
\reference{anm}Abraham, R.G. \& Merrifield, M.R. 2000, \aj, 120, 2835
\reference{a94}Abraham, R.G., Valdes, F., Yee, H.K.C. \& van den Bergh, S. 
1994, \apj, 432, 1
\reference{bqp}Berlind, A.A., Quillen, A.C., Pogge, R.W. \& Sellgren, K. 1997,
\aj, 114, 107
\reference{bea}Block, D.L., Bertin, G., Stockton, A., Grosb\o l, P., Moorwood,
A.F.M. \& Peletier, R.F. 1994, \aap, 288, 365
\reference{bnp}Block, D.L. \& Puerari, I. 1999, \aap, 342, 627
\reference{bw}Block, D.L. \& Wainscoat, R.J. 1991, Nature, 353, 48
\reference{rng}Buta, R. 1986, \apjs, 61, 609
\reference{rn2}Buta, R. 1995, \apjs, 96, 39
\reference{bnb}Buta, R. \& Block, D.L. 2001, \apj, 550, 243
\reference{cbj}Conselice, C.J., Bershady, M.A. \& Jangren, A. 2000, \apj, 529,
886
\reference{d93}DePoy, D.L., Atwood, B., Byard, P.L., Frogel, J. \& O'Brien, 
T.P. 1993, in Infrared Detectors and Instrumentation, Proc.~SPIE Vol.~1946 
A.M.~Fowler, ed.~p.~667
\reference{dvd}de Vaucouleurs, G. 1959, Handb.~der Physik, 53, 275
\reference{rc2}de Vaucouleurs, G., de Vaucouleurs A. \& Corwin, H.G., Jr. 1976,
Second Reference Catalogue of Bright Galaxies (University of Texas Press:  
Austin)
\reference{rc3}de Vaucouleurs, G., de Vaucouleurs A., Corwin, H.G., Jr., Buta, 
R.J., Paturel, G. \& Fouqu\'e, P. 1991, Third Reference Catalogue of Bright 
Galaxies (Springer-Verlag:  New York) (RC3)
\reference{ee2}Elmegreen, D.M. \& Elmegreen, B.G. 1982, \mnras, 201, 1021
\reference{ene}Elmegreen, B.G. \& Elmegreen, D.M. 1985, \apj, 288, 438
\reference{ees}Elmegreen, D.M. \& Elmegreen, B.G. 1987, \apj, 314, 3
\reference{ane}Elmegreen, D.M., Elmegreen, B.G., Frogel, J.A., Eskridge, P.B.,
Pogge, R.W., Gallagher, A. \& Iams, J. 2002, \aj, in press (astro-ph/0205105)
\reference{bar}Eskridge, P.B., Frogel, J.A., Pogge, R.W., Quillen, A.C., 
Davies, R.L., DePoy, D.L., Houdashelt, M.L., Kuchinski, L.E., Ram\'{\i}rez, 
S.V., Sellgren, K., Terndrup, D.M. \& Tiede, G.P. 2000, \aj, 119, 536
\reference{ezo}Eskridge, P.B., Taylor, V.A., Windhorst, R.A., Odewahn, S.C.,
Chiarenza, C.A.T., Conselice, C.J., de Grijs, R., Matthews, L.D., O'Connell,
R.W., Frogel, J.A. \& Gallagher, J.S. 2001, \baas, 33, 1379
\reference{fgt}Frei, Z., Guhathakurta, P., Gunn, J.E. \& Tyson, J.A. 1996, \aj,
111, 174
\reference{hs}Hackwell, J.A. \& Schweizer, F. 1983, \apj, 265, 643
\reference{tnf}Hubble, E.P. 1936, The Realm of the Nebulae (Yale University
Press:  New Haven)
\reference{kea}Kuchinski, L.E., Terndrup, D.M., Gordon, K.D. \& Witt, A.N. 
1998, \aj, 115, 1438
\reference{knt}Kuchinski, L.E. \& Terndrup, D.M. 1996, \aj, 111, 1073
\reference{bio}Mayr, E. 1942, Systematics and the Origin of Species (Columbia
University Press:  New York).
\reference{wwm}Morgan, W.W. 1958, \pasp, 70, 364
\reference{mkw}Morgan, W.W., Kayser, S. \& White, R.A. 1975, \apj, 199, 545
\reference{nea}Naim, A., Lahav, O., Buta, R.J., Corwin, H.G., Jr., de 
Vaucouleurs, G., Dressler, A., Huchra, J.P., van den Bergh, S., Raychaudhury, 
S., Sodre, L., Jr. \& Storrie-Lombardi, M.C. 1995, \mnras, 274, 1107
\reference{nrg} Naim, A., Ratnatunga, K.U. \& Griffiths, R.E. 1997, \apjs, 111,
357
\reference{onp}Odewahn, S.C., Cohen, S.H., Windhorst, R.A. \& Philip, N.S. 
2002, \apj, 568, 539
\reference{oea}Odewahn, S.C., Windhorst, R.A., Driver, S.P. \& Keel, W.C. 
1996, \apj, 472, L13
\reference{old}Parsons, W., Lord Rosse, 1850, \mnras, 10, 21
\reference{paq}Patsis, P.A., Athanassoula, E. \& Quillen, A.C. 1997, \apj, 483,
731
\reference{p98}Pogge, R.W., DePoy, D.L., Atwood, B., O'Brien, T.P., Byard, 
P.L., Martini, P., Stephens, A., Gatley, I., Merrill, K.M., Vrba, F.J. \& 
Henden, A.A. 1998, in Infrared Astronomical Instrumentation, Proc.~SPIE 
Vol.~3354, A.M.~Fowler, ed., p.414
\reference{pea}Puerari, I., Block, D.L., Elmegreen, B., Frogel, J.A. \& 
Eskridge, P.B. 2000, \aap, 359, 932
\reference{qfg}Quillen, A.C., Frogel, J.A. \& Gonzalez, R.A. 1994, \apj, 437, 
162
\reference{qnd}Quillen, A.C. \& Frogel, J.A. 1997, \apj, 487, 603
\reference{qkd}Quillen, A.C., Frogel, J.A., Kenney, J.D.P, Pogge, R.W. \& 
DePoy, D.L. 1995, \apj, 441, 549
\reference{qea}Quillen, A.C., Kuchinski, L.E., Frogel, J.A. \& DePoy, D.L., 
1997, \apj, 481, 179
\reference{sng}Regan, M.W., Thornley, M.D., Helfer, T.T., Sheth, K., Wong, T.,
Vogel, S.N., Blitz, L. \& Bock, D.C.-J. 2001, \apj, 561, 218
\reference{r98} Rhoads, J.E. 1998, \aj, 115, 472
\reference{hag}Sandage, A. 1961, The Hubble Atlas of Galaxies (Carnegie 
Institute of Washington Press:  Washington)
\reference{cag}Sandage, A. \& Bedke, J. 1994, The Carnegie Atlas of Galaxies
(Carnegie Institute of Washington:  Washington DC)
\reference{sch}Schweizer, F. 1976, \apjs, 31, 313
\reference{tea}Terndrup, D.M., Davies, R.L., Frogel, J.A., DePoy, D.L. \& 
Wells, L.A., 1994, \apj, 432, 518
\reference{hfd}Thronson, H.A., Jr., Hereld, M., Majewski, S., Greenhouse, M.,
Johnson, P., Spillar, E., Woodward, C.E., Harper, D.A. \& Rauscher, B.J. 1989,
\apj, 343, 158
\reference{vea}van den Bergh, S., Abraham, R.G., Whyte, L.C., Merrifield, M.R.,
Eskridge, P.B., Frogel, J.A. \& Pogge, R.W. 2002, \aj, in press
(astro-ph/0202444)
\reference{v76}van den Bergh, S. 1976, \apj, 206, 883
\reference{xes}Whitmore, B.C. \& Bell, M. 1988, \apj, 324, 741
\reference{wyt}Whyte, L.F., Abraham, R.G., Merrifield, M.R., Eskridge, P.B.,
Frogel, J.A. \& Pogge, R.W. \mnras, submitted
\reference{fzw}Zwicky, F. 1955, \pasp, 67, 232
\end{references}
\end{document}